\documentclass[prd,aps,superscriptaddress,10pt,nobibnotes,notitlepage,twocolumn,longbibliography,nofootinbib,preprintnumbers,dvipsnames]{revtex4-2}

\usepackage{cool}
\usepackage[normalem]{ulem}
\usepackage{sidecap}
\usepackage[latin1]{inputenc}
\usepackage{graphicx}
\usepackage{float}
\usepackage{latexsym}
\usepackage{graphicx}
\usepackage{amssymb}
\usepackage{amsmath,mathtools}
\usepackage{amsfonts}
\usepackage{tikz} 
\usepackage{ragged2e}
\usepackage{stackengine}
\usepackage{scalerel}

\newcommand{\lowersub}{\smash[b]{\vphantom{\bigg|}}}

\usepackage[colorlinks=true,citecolor=blue,hyperfootnotes=false]{hyperref}

\usepackage{dcolumn}
\usepackage{textcomp}
\usepackage{xfrac}
\usepackage{slashed}
\usepackage{multirow}

\def\be{\begin{equation}}
\def\ee{\end{equation}}
\def\figs/B{B}
\def\bea{\begin{eqnarray}}
\def\eea{\end{eqnarray}}
\def\bg{\begin{eqnarray}}
\def\nd{\end{eqnarray}}
\def\sin{{\rm sin}}

\def\beq{\begin{equation}}
\def\eeq{\end{equation}}

\newcommand{\intd}{\int {\rm d}}
\newcommand{\sint}[3]{\hspace{- #3 em} \underset{#1}{\overset{#2}{\int}}\hspace{- #3 em}}

\usepackage{graphicx}
\usepackage{dcolumn}
\usepackage{bm}
\usepackage{hyperref}
\usetikzlibrary{decorations.pathmorphing}
\tikzset{snake it/.style={decorate, decoration=snake}}

\begin{document}

\preprint{MIT-CTP/5672}

\title{Toward quantum tunneling from excited states: \\
Recovering imaginary-time instantons from a real-time analysis}

\author{Thomas Steingasser}
 \email{tstngssr@mit.edu}
\affiliation{%
Department of Physics, Massachusetts Institute of Technology, Cambridge, MA 02139, USA }
\affiliation{Black Hole Initiative at Harvard University, 20 Garden Street, Cambridge, MA 02138, USA
}%

\author{David I. Kaiser}
 \email{dikaiser@mit.edu}
\affiliation{%
Department of Physics, Massachusetts Institute of Technology, Cambridge, MA 02139, USA
}%

\date{\today}

\begin{abstract}

We revisit the path integral description of quantum tunneling and lay the groundwork for its generalization to excites states through real-time path integral techniques. For clarity, we focus on the simple toy model of a point particle in a double-well potential, for which we perform all steps explicitly. Instead of performing the familiar Wick rotation from physical to imaginary time---which is inconsistent with the requisite boundary conditions when treating tunneling from states other than the false vacuum---we regularize the path integral by adding an infinitesimal complex contribution to the Hamiltonian, while keeping time strictly real. We find that this gives rise to a complex stationary-phase solution, in agreement with recent insights from Picard-Lefshitz theory. We then show that there exists a class of analytic solutions for the corresponding equations of motion, which can be made to match the appropriate boundary conditions in the physically relevant limits of a vanishing regulator and an infinite physical time. We provide a detailed discussion of this non-trivial limit. We find that, for systems without an explicit time-dependence, our approach reproduces the picture of an instanton-like solution defined on a finite Euclidean-time interval. Lastly, we discuss the generalization of our approach to broader classes of systems, for which it serves as a reliable framework for high-precision calculations.

\end{abstract}

\maketitle
\section{Introduction}

Tunneling is a universal feature of all quantum theories~\cite{Coleman:1977py,Callan:1977pt,Kobzarev:1974cp,Devoto:2022qen}. Despite being textbook knowledge for decades, it also remains a subject of active research, driven by recent progress in the understanding of its path-integral formulation. For the purpose of this article, two of these developments are of particular interest. First, the \textit{direct approach} to tunneling brought forward recently in Refs.~\cite{Andreassen:2016cff,Andreassen:2016cvx,Andreassen:2017rzq} and extended in Refs.~\cite{Khoury:2021zao,Steingasser:2022yqx,Chauhan:2023pur} has clarified several important properties
of the tunneling-rate calculation. This approach relies on the use of \textit{instantons}, which are solutions of the equations of motion in imaginary time. In this picture, the imaginary time arises through a simple Wick rotation from physical time, allowing for a straightforward identification of the infinitely long imaginary time interval with the limit of an infinitely long physical time, relative to the typical time-scales of the system. This limit has a strong physical motivation and it is necessary to avoid so-called \textit{sloshing} effects, as described in Refs.~\cite{Andreassen:2016cff,Andreassen:2016cvx}. Meanwhile, on a practical level, the Wick rotation is necessary to realize the instanton's boundary conditions. Using a point particle for concreteness, the instanton can be understood as its motion in the inverted potential, starting at rest from the false vacuum at the initial (imaginary) time. 

A straightforward generalization of this simple picture would suggest that the decay rate out of an excited state should be determined by the Euclidean action of a suitable instanton solution connecting the initial position of interest with the emergence point on the other side of the potential barrier. As the initial state is no longer a local extremum, this motion can be expected to occur within some finite imaginary-time interval, whose duration should be of the order of the potential's natural time-scale. In other words, for tunneling out of a state different from the false vacuum, one might expect that the imaginary-time interval of interest would be determined by the form of the potential and the initial state itself. This picture agrees with the result obtained through the WKB approximation, in which the imaginary time appears as a formal tool rather than a representation of the physical time~\cite{Callan:1977pt,Coleman:1977py,LandauLifshitzQM,Griffiths}. 

In the direct approach, however, the imaginary time is usually linked to the real, physical time, from which it arises through a Wick rotation. This would seem to suggest that the length of the Euclidean-time interval is determined by the physical time that has passed since the particle started out from its initial state. Taking seriously this interpretation, our previous observations would then imply that the instanton only exists for \textit{one} unique (physical) time, of order of the system's natural time-scale. This conclusion not only appears counterintuitive on its own, but also disagrees with the necessity of a physical time that is significantly longer than the system's natural time-scale, as found in Refs.~\cite{Andreassen:2016cff,Andreassen:2016cvx}. 

Clearly, the natural candidate for the origin of this issue is the Wick rotation, which usually implicitly relies on the assumption that the system approaches a minimum of the potential at asymptotic physical times. Thus, when interested in the tunneling rate out of a general state, one is forced to work in real time. To make this manifest, as we demonstrate here, one may instead regularize the path integral by introducing an infinitesimal imaginary 
part of the Hamiltonian. For time-independent systems, this is equivalent to performing an infinitesimal Wick rotation and eventually taking the appropriate limit.\footnote{The interpretation of a complex energy arises naturally when deriving the regularization from the properties of a physical initial state (see, e.g., Refs.~\cite{SchwartzQFT,Steingasser:2023gde}). The conceptual subtleties linked to the common step of assigning the imaginary part to the time variable are illustrated, e.g., in Ref.~\cite{Kaya:2018jdo}.} However, this equivalence is broken for systems with an explicit time-dependence, for which the usual Wick rotation is even more subtle. 

Because our analysis relies entirely on real, physical time, it is ideally suited for the investigation of such systems. Most important, we show that our approach can be used to properly derive the length of the imaginary-time interval on which the instanton is to be defined. This feature is crucial for the investigation of more complicated systems, such as near evaporating black holes, for which the precise imaginary-time description of the tunneling process remains controversial~\cite{Hiscock:1987hn,Berezin:1987ea,Arnold:1989cq,Gregory:2013hja,Burda:2015isa,Burda:2015yfa,Burda:2016mou,Gorbunov:2017fhq,Shkerin:2021zbf,Cheung:2013sxa,Tetradis:2016vqb,Canko:2017ebb,Mukaida:2017bgd,Kohri:2017ybt,Hayashi:2020ocn}. By providing a simple tool with which to derive the decay rate as well as the properties of the relevant Euclideanized spacetime, we aim to take the first step towards providing clarity regarding this important matter. 

While we perform our calculations exclusively in real time, the familiar leading-order results imply that, when dealing with systems with no explicit time-dependence, we should ultimately be able to express our results in terms of an (imaginary-time) instanton. Making the connection between these two perspectives requires an understanding of instanton-like solutions on time-contours with an arbitrary Wick-rotation angle in the complex-time plane. For tunneling out of a false vacuum, there has recently been significant progress in this direction within the framework of Picard-Lefshetz theory~\cite{Witten:2010cx,Tanizaki:2014xba,Cherman:2014sba,Dunne:2015eaa,Bramberger:2016yog,Michel:2019nwa,Mou:2019gyl,Hertzberg:2019wgx,Ai:2019fri,Hayashi:2021kro,Nishimura:2023dky}. In particular, it has been shown in great detail in Ref.~\cite{Ai:2019fri} that the decay rate of the false vacuum is independent of the Wick-rotation angle. 

Our computations are consistent with this result, while further clarifying that the dynamics along the complex-time contour constitute a ``shearing'' rather than a rotation: the convergence of the solution is controlled primarily by the change of the complex-time variable along the imaginary-time direction. In addition, we find that the limit of a vanishing angle between the contour and the real-time axis is, in fact, \textit{not} independent of the physical time. In the case of an infinite imaginary time, this makes no practical difference, since any contour in the complex-time plane with a finite angle relative to the real-time axis will eventually lead to a sufficiently large imaginary part to ensure convergence. For instantons defined on a finite imaginary-time interval, on the other hand, this implies that any contour along a finite angle will contain an infinite number of instantons in the infinite-time limit, reproducing the problem already present for the purely imaginary contour. We find that this problem can be cured by restricting oneself to a particular subset of possible complex-time contours. This step can be easily justified through our interpretation of the direct approach, and provides a strong justification for the restriction of contours to the so-called Lefshetz thimbles in the Picard-Lefshetz theory. 

In Sec.~\ref{sec:ComplexTime}, we review the emergence of imaginary time in the leading-order analysis of quantum tunneling, summarize the recently proposed direct approach, and provide a first sketch of our results. Next, in Sec.~\ref{sec:FirstPrinciples}, we review in more detail the derivation of the decay rate in the direct approach and generalize it to an arbitrary initial state. We then discuss the regularization scheme we use to evaluate the resulting path integrals in Sec.~\ref{sec:RegularizedDynamics}, where we also provide a simple but helpful interpretation of the corresponding saddle-point approximation in terms of interacting, non-relativistic point particles. 

In Sec.~\ref{sec:Steadyon}, we analyze the equivalent of the instanton in our real-time framework, which we dub a \textit{steadyon}. We discuss in particular the ability of these solutions to reproduce the boundary conditions of the path integral of interest, and show how they can be used to deduce the tunneling rate in the combined limit of a vanishing regulator and an infinite (real) physical time. Crucially, we demonstrate how our approach can be used to derive the imaginary-time picture for systems without an explicit time-dependence, to leading order. An important feature of the direct approach, as we show, is that it enables the computation of the appropriate normalization factor for the path integral, in addition to identifying the terms associated with the instanton. In Sec.~\ref{sec:Drag}, we show how our approach can be used to evaluate this expression, and illustrate how each relevant term can be understood in terms of imaginary-time dynamics. Finally, in Sec.~\ref{sec:Conclusion}, we review our work with a special emphasis on its generalization to more complicated systems, such as field theory or systems with an explicit time-dependence.

\section{Complex time in quantum tunneling}\label{sec:ComplexTime}

For simplicity, we consider a point particle of mass $m$ and energy $E$ trapped by some potential barrier in one spatial dimension. Assuming the tunneling rate to be proportional to the transmission coefficient, it is straightforward to use the WKB approximation to show that the tunneling rate takes the general form~\cite{Coleman:1977py,Callan:1977pt,LandauLifshitzQM,Griffiths}
\begin{align}
    \Gamma =  A e^{- B} \left[ 1 + {\cal O} (\hbar) \right],
\end{align}
where $B$ is the WKB factor
\begin{align}
    B=  2 \int_{x_i}^{x_s} {\rm d}x \left[2 m (V(x)-E) \right]^{\frac{1}{2}}.
\end{align}
Here $x_i$ is the classical turning point of the particle, while $x_s$ is the point on the other side of the barrier for which the particle has the same energy.

In order to obtain a simple interpretation of this result, it was famously proposed by the authors of Refs.~\cite{Coleman:1977py,Callan:1977pt} to consider the motion of the same point particle in imaginary time $t \to - i \tau $ along the classical path $\bar{x}_I$ that connects the points $x_i$ and $x_s$, with no turning points. The energy of the particle is then given by
\begin{align}
     E = V (\bar{x}_{\rm I}) - \frac{1}{2}m\dot{\bar{x}}_{\rm I}^2 (\tau) ,
\end{align}
where dots denote derivatives with respect to $\tau$. It is straightforward to show that the WKB factor $B$ is fully determined by this solution:
\begin{align}
    B= &  \, 2 \int_{x_i}^{x_s} {\rm d}x \left[2 m (V(x)-E) \right]^{\frac{1}{2}}  \nonumber \\
    =& \, 2 m  \int_{x_i}^{x_s} {\rm d}x
    \  \dot{\bar{x}}_I (\tau (x)) = 2 m \int_{- \Delta \tau_I /2}^{0} {\rm d}\tau \ \dot{\bar{x}}_I^2  \nonumber \\ 
    =& \int_{- \Delta \tau_I /2 }^{\Delta \tau_I /2} {\rm d}\tau \ \left[ \frac{m}{2} \dot{\bar{x}}_I^2 + V (\bar{x}_I (\tau)) -E \right]  \nonumber \\ 
    =& \, S_E [\bar{x}_I] - E \cdot \Delta \tau_I , \label{eq:B}
\end{align}
where $\Delta \tau/2$ is defined as the (Euclidean) time necessary for the particle to move from $x_i$ to $x_s$. In the second line, the invertibility of $\bar{x}_I (\tau)$ has been used. 

Eq.~\eqref{eq:B} implies that the exponent of the decay rate is, to leading order and up to the term $E \cdot \Delta \tau_I$, given by the Euclidean action of the solution $\bar{x}_I$. The latter can be recognized as the familiar instanton, while the $E \cdot \Delta \tau_I$ term can be linked to the fact that the initial and final states are resonance states~\cite{Liang:1992ms,Liang:1994xn,Liang:1994qr,Muller-Kirsten:2001jhw}. For our purpose, the most interesting insight gained from this discussion is that the (finite) length of the Euclidean time interval $\Delta \tau_I$ is by construction such that a suitable instanton connecting $x_i$ and $x_s$ can exist. 

Whereas this approach successfully captures the leading-order tunneling rate in many situations, in particular out of the false vacuum, its range of application is limited due to its reliance on the WKB approximation. For example, on a conceptual level, there is no known procedure to extend this approach to incorporate loop corrections or dynamical effects. Even at leading order, the WKB approach breaks down in regions for which $V(x) \sim E$. This includes, in particular, tunneling from an initial state that is relatively close to the top of the potential barrier~\cite{LandauLifshitzQM,Griffiths}. This problem is of particular significance for tunneling out of relatively flat minima. Potentials with such features have recently gained interest in the context of new approaches to beyond-the-Standard Model (BSM) model building~\cite{Steingasser:2023ugv,Benevedes:2025qwt}.

A more rigorous formulation of quantum tunneling can be obtained in terms of path integrals. Following Refs.~\cite{Andreassen:2016cff,Andreassen:2016cvx}, the decay rate can be brought to the form
\begin{align}
    \Gamma (T)=&  \frac{1}{Z} \int_{x_1 (0)=x_i}^{x_1 (T)=x_s} {\mathcal D} x_1 \ e^{i S[x_1] } \delta \left( F_{x_s}[x_1] - T \right)  \nonumber \\ 
    &\quad\quad \times \left(\int_{x_2 (0)=x_i}^{x_2 (T)=x_s} {\mathcal D} x_2 \  e^{ i S [x_2] } \right)^* + c.c. ,
    \label{eq:Gamma1}
\end{align}
where $T$ denotes the physical time at which the tunneling rate is considered, and the functional $F_{x_s} [x_1]$ identifies the time at which the path $x_1$ first crosses $x_s$. (We provide a complete derivation of this result in Sec.~\ref{sec:FirstPrinciples}.) In order to arrive at the familiar imaginary-time picture, the authors of Refs.~\cite{Andreassen:2016cff,Andreassen:2016cvx} then propose to perform a Wick rotation, $T \to  i \tau$. An alternative derivation of the decay rate, leading to an equivalent result, has been performed in Ref.~\cite{Ai:2019fri}. The latter also relies on a Wick rotation to reproduce the familiar imaginary-time results, while further considering (incomplete) Wick rotations by some finite angle in the complex-time plane, demonstrating that the final results are independent of the rotation angle.

When considering tunneling out of the false vacuum, these results are perfectly consistent with the one obtained from the WKB approximation. In particular, they successfully reproduce the infinite imaginary-time interval of the instanton in the limit of large physical times $T \to \infty$ motivated in Refs.~\cite{Andreassen:2016cff,Andreassen:2016cvx}. However, it is easy to see that this procedure cannot be used to describe tunneling out of a more general state, for which the Wick rotation would seem to require an infinitely long imaginary-time interval, given the infinite-$T$ limit. Comparing with Eq.~(\ref{eq:B}), we see that taking the limit $\Delta \tau_I \rightarrow \infty$ is inconsistent with the WKB approximation for a resonance state.

In addition, performing the usual Wick rotation and taking the limit $\Delta \tau_I \rightarrow \infty$ would prevent us from using the saddle-point approximation to evaluate the path integrals in the first place. The initial condition implies that the instanton begins and ends its motion in points for which $V^{\prime} \neq 0$. This, in turn, implies that the particle's periodic motion occurs within some finite imaginary time, of order of the system's dynamical time-scale~\cite{Liang:1992ms,Liang:1994xn,Liang:1994qr,Muller-Kirsten:2001jhw}. In other words, if one were to perform the usual Wick rotation to imaginary time, the tunneling could only be described through the saddle-point approximation for some unique (real) time $T$, and, in particular, \textit{not} in the physically relevant limit $T \to \infty$.\footnote{This issue cannot be avoided by taking into account the possibility of \textit{multi-instantons}, as the limit of an infinite imaginary time would correspond to an infinite number of instantons, and hence, an infinite Euclidean action.}

\begin{figure*}[t!]
    \centering
    \includegraphics[width=0.32\textwidth]{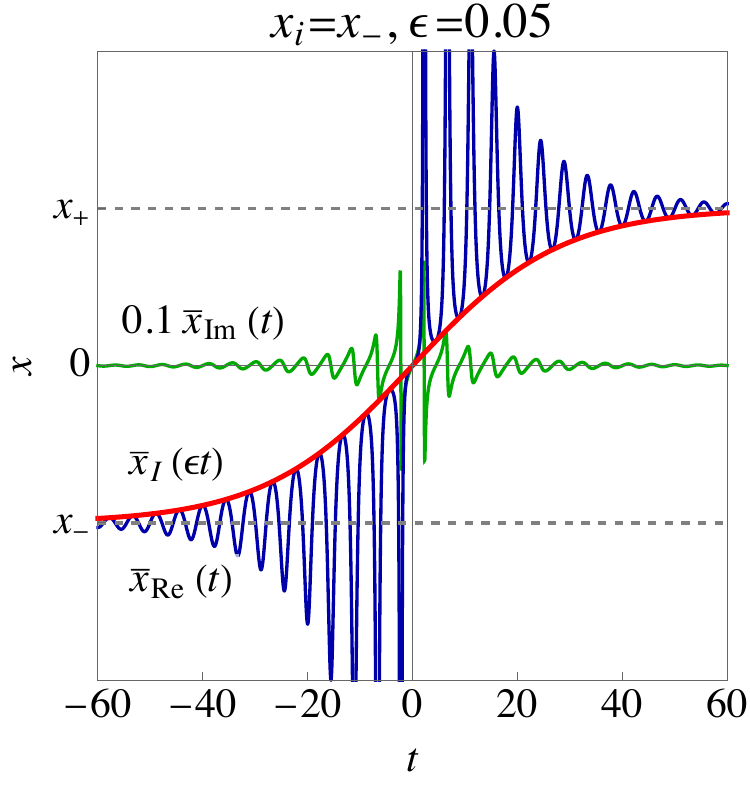}
    \includegraphics[width=0.32\textwidth]{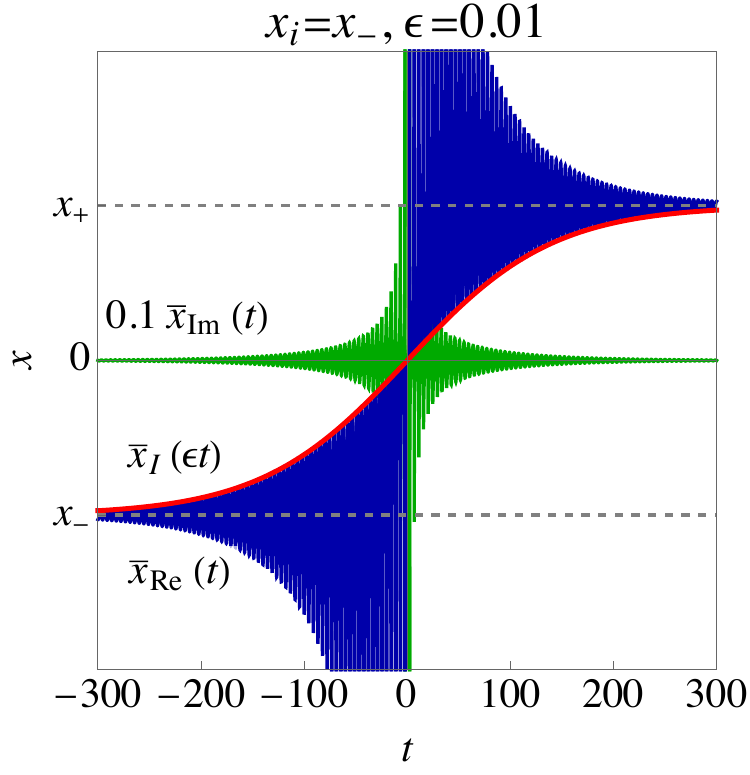}
    \includegraphics[width=0.32\textwidth]{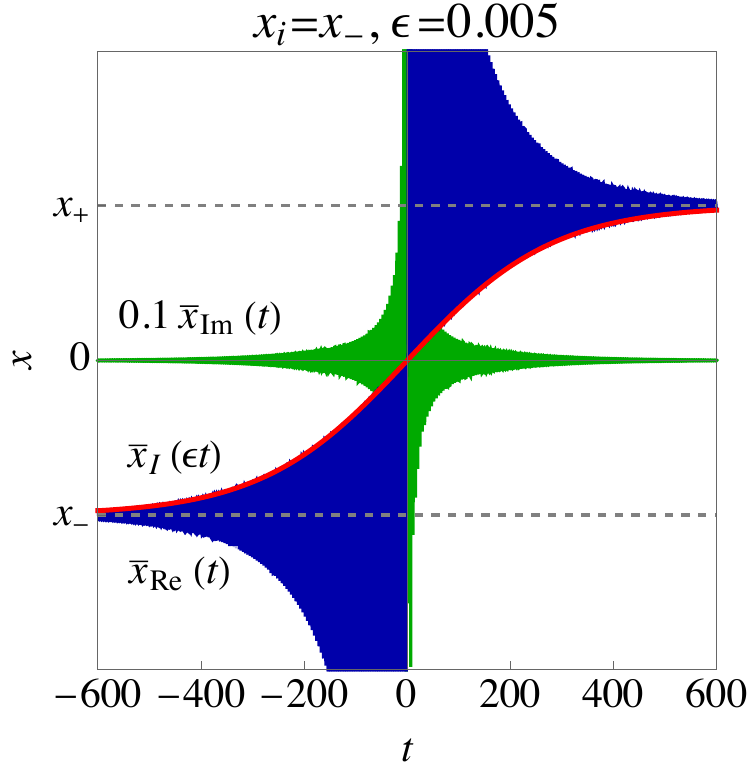}
    \caption{The well-known kink solution describing tunneling out of the false vacuum for a particle in a double-well potential (with vacua at $x_\pm$) after a finite Wick rotation $\tau \to i (1-i \epsilon) t$. By considering arbitrary rotation angles, the instanton solution $\bar{x}_I (\tau)$ (red curve) becomes complex, and its convergence toward the desired initial and final values is controlled by the combination $\epsilon \cdot t$. It was shown in Ref.~\cite{Ai:2019fri} that the decay rate obtained using this class of solutions agrees with the imaginary-time result.}
    \label{fig:FVCI}
\end{figure*}

To understand the origin of this contradiction, we note that, on a qualitative level, the case of tunneling from the false vacuum is unique due to the asymptotic behavior of the corresponding instanton: it converges towards the false vacuum, whereas a generic solution would undergo an infinite, periodic motion. A simple way around the complications arising from the latter case would be to avoid the Wick rotation entirely and evaluate the path integral in real time. This idea has been pursued in Ref.~\cite{Ai:2019fri}, which analyzes tunneling out of a false vacuum by performing a Wick rotation $t \to e^{- i \epsilon} t$ and taking the limit $\epsilon \to 0$. For the special case of tunneling out of the false vacuum, and thus considering an infinitely long time interval, these authors find that the decay rate is independent of $\epsilon$.

This result hinges, however, on the infinite length of the time interval under consideration. To understand this dependence, we may consider the tunneling of a point particle in a potential with two (nearly) degenerate vacua $x_{\pm}$. In imaginary time, the instanton describing this process to leading order is the well-known \textit{kink}, whose value at $\tau \to - \infty $ (or, equivalently, $ T \to - \infty $) describes the initial state of the system. Along a more general complex-time contour, the solution itself becomes a complex function, 
\begin{equation}
\bar{x}_I (e^{i \epsilon} t)= \bar{x}_{\rm Re}(t) + i  \bar{x}_{\rm Im}(t) . 
\label{eq:barxI}
\end{equation}
In the limit $\epsilon \to 0$, the time evolution of $\bar{x}_{\rm Re}$ and $\bar{x}_{\rm Im}$ can be decomposed into an oscillatory motion with the system's typical, classical frequency as well as a time evolution of the amplitude, which is responsible for the convergence of these functions to their desired initial values. See Fig.~\ref{fig:FVCI}.

Most important for our purpose, we observe that the evolution of the amplitude of the real part of the complex-valued instanton reproduces the behavior of the imaginary-time instanton under the identification $\tau = \epsilon t$. In simple terms, this suggests that the transition to a complex-time argument involves a {\it projection} rather than a rotation, see Fig.~\ref{fig:FVCI}. For tunneling out of the false vacuum, this still ensures the convergence of the complex-time solution, as the limit $T \to \infty$ amounts to $\tau \to \infty$ for \textit{any} non-vanishing $\epsilon$. As we will see, the generalization of this observation will turn out to be crucial for the explanation of the finite imaginary-time support of the instanton corresponding to tunneling out of a state other than the false vacuum. 

This insight, which represents one of the main results of this article, is discussed in detail in Sec.~\ref{sec:Steadyon} and summarized schematically in Fig.~\ref{fig:DeltaTauOrigin}. In essence, we argue that the Wick rotation angle $\epsilon$ is to be understood as a regularization parameter, and hence it must be taken to zero. As we also take the limit $T \to \infty$, it becomes possible to take these limits in such a way that the combination $\epsilon \cdot T$ remains constant, ensuring the existence of the complex-valued instanton solution. This solution is, however, vastly different from the more familiar imaginary-time instanton. First, taking the limit $\epsilon \to 0$ amounts to rotating the time contour approximately onto the real-time axis. Second, due to the limit $T \to \infty$, both the real and imaginary parts of the complex instanton solutions become highly oscillatory functions, leading to a divergent real part of the (now complex) action. 

In Secs.~\ref{sec:Steadyon} and~\ref{sec:Drag}, we demonstrate explicitly for our toy model how these configurations lead to a sensible result for the decay rate, which agrees to leading order with the naive expectation based on the imaginary-time picture. We find, in particular, that the exponent of the decay rate is entirely determined by the well-behaved imaginary part of the action, which in the combined limit $\epsilon \to 0 ,\, T \to \infty$ converges to the Euclidean action of the corresponding imaginary-time instanton.

\section{Tunneling rate from first principles}\label{sec:FirstPrinciples}

\subsection{Defining the tunneling rate}

\begin{figure}[h!]
    \centering
    \includegraphics[width=0.48\textwidth]{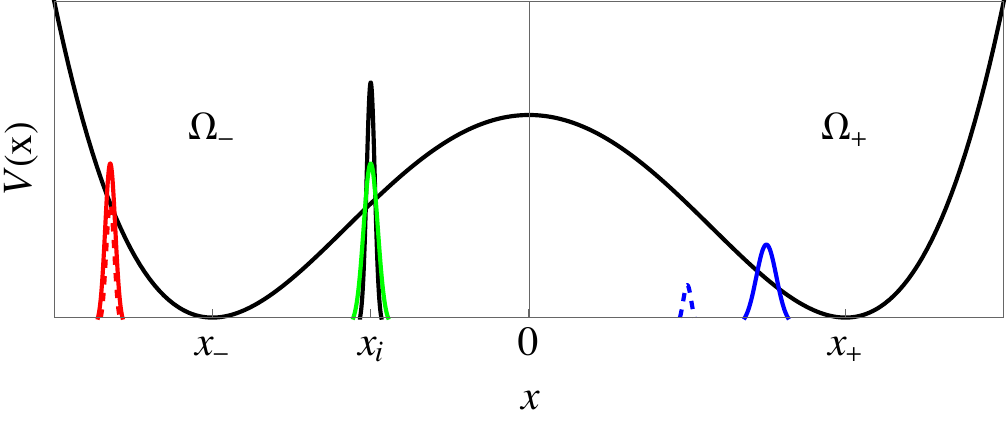}
    \caption{A visualization of the ansatz of~Eq.~\eqref{Gammadef}. At $t=0$, the point particle is localized at $x_i$ (black curve). Starting from this position, it will either tunnel to region $\Omega_+$ with an exponentially suppressed probability (solid blue peak in region $\Omega_+)$, or move to the left. Following the latter path, the particle will reach a turning point (solid red peak within $\Omega_-$) after some time $\Delta t_R/2$. The particle would then oscillate back toward $x_i$, returning with some diminished probability (green curve). Then the process will repeat, with correspondingly smaller probabilities both to tunnel into $\Omega_+$ (dashed blue curve) or to oscillate to the left within $\Omega_-$ (dashed red curve). }
    \label{fig:Potential}
\end{figure}

We consider a point particle in a potential with two vacua $x_{\pm}$, surrounded by two basins $\Omega_{\pm}$. See Fig.~\ref{fig:Potential}. In the most general case, the initial state at $t=0$ of such a particle can be described through a density matrix $\rho = \intd x_i \intd x_j\ \rho_{ij}  | x_i \rangle \langle x_j |$. For a pure state with wave function $\psi (x)$, the coefficient $\rho_{ij}$ takes the simple form $\rho_{ij}=\psi (x_i) \psi^* (x_j)$. For the remainder of this article, we will restrict ourselves to the idealized case of a highly localized (but on its own unphysical) initial state, $\rho = | x_i \rangle \langle x_i |$. For our present analysis, we consider our simplified choice because it allows for a clear understanding of many conceptual aspects of our computation. Moreover, any density matrix can be expressed in terms of matrix elements $| x_i \rangle \langle x_j |$. As argued in Ref.~\cite{Steingasser:2023gde}, symmetry suggests that for generic systems the dominant contribution will arise from stationary phase solutions with $x_i = x_j$. Whenever this is the case, this allows to represent the tunneling rate out of \textit{any} initial state as the superposition of the rates out of individual position eigenstates, each of which can be calculated using our results---in particular, excited states. However, even in cases in which non-diagonal elements of $\rho_{ij}$ should become important we anticipate that it is straightforward to generalize our results accordingly. Crucially, the state we consider is of course itself \textit{not} a physical state, but rather serves the purpose of making transparent the semi-classical structure of the relevant class of path integrals. For an example of an actual physical state, consider Ref.~\cite{Steingasser:2023gde}

We may now anticipate the relevant behavior of the system using the interpretation of a path integral as a superposition of all possible paths, dominated by the particle's classical trajectory. First, we consider a particle at rest at the position $x_i$ at $t=0$. Classically, such a particle can then be expected to perform an oscillatory motion within the basin $\Omega_{-}$. We can furthermore expect that it is most likely for this particle to tunnel into the basin $\Omega_{+}$ at the times when it is close to the turning point $x_i$. During these times, the total probability transfer can be expected to be proportional to the total probability within $\Omega_{-}$, $\Delta P_{\Omega_{-}} \sim \textcolor{red}{-}\mathcal{T} \cdot P_{\Omega_{-}}$, with the usual transmission coefficient $\mathcal{T}$ usually calculated through the WKB method. Thus, \textit{during these time intervals}, we expect the probability to find the particle within the basin $\Omega_{-}$ to evolve with time as
\begin{equation}\label{Gammadef}
   P_{\Omega_{-}}(T) = P_{\Omega_{-}}(0) \cdot e^{- \Gamma  T},
\end{equation}
where $\Gamma$ is the decay rate we ultimately wish to compute, and we have assumed that $\Gamma$ is approximately constant in the regime of interest, as is the case, e.g., in the limit $T \to \infty$. It is straightforward to solve Eq.~\eqref{Gammadef} for $\Gamma$, yielding
\begin{equation}
    \Gamma (T) = - \frac{1}{P_{\Omega_{-}}(T)} \frac{\text{d}}{\text{d} T}P_{\Omega_{-}}(T)=\frac{1}{P_{\Omega_{-}}(T)} \frac{\text{d}}{\text{d} t}P_{\Omega_{+}}(T), \label{Gammadef2}
\end{equation}
where $\Omega_{+}$ is the region in position space into which the particle tunnels. Henceforth we adopt the abbreviation $Z (T) \equiv P_{\Omega_{-}}(T)$. For the remainder of this article, we will use the quantity $\Gamma$ as the \textit{definition} of the tunneling rate.

The probability to recover the point particle within \textit{some} region $\Omega$ (such as $\Omega_{-}$) at some later time $t=T$ is simply given by~\cite{Shkerin:2021zbf}
\begin{align}
    P_{\Omega} (T)=& \text{Tr}_\Omega\left[ \rho \right] (T) = \int_{\Omega} d x \langle x,T | x_i\rangle \langle x_i | x,T \rangle . \label{fund}
\end{align}
Here and throughout the remainder of our article, all states with no explicit time label are to be understood as defined at $t=0$, e.g. $| x_i\rangle \equiv | x_i ,0\rangle$. 

An important subtlety of this ansatz is that we have arrived at it by considering a path in which the particle starts with vanishing initial momentum. It is, however, straightforward to see that the above behavior can be found for any initial momentum, albeit with different times $\Delta t_R$. Conversely, for any given time $T$, the loss of probability within $\Omega_-$ is of the form of Eq.~\eqref{Gammadef}, where $\Gamma$ is determined by the contribution representing a different initial momentum. We will revisit this matter in Sec.~\ref{sec:NPSteadyonSolution}, where we will show how this picture emerges in a natural way from our calculation.

\subsection{Path Integral representation}

The last factor in Eq.~\eqref{Gammadef2} can be represented through the point particle's propagator $D_F (x_i, t_i | x_f, t_f)$ as
\begin{align}
    P_{\Omega_{+}} = & \int_{\Omega_{+}} {\rm d} x_f \ D_F( x_f , T | x_i ) D_F(x_i| x_f,T  ).
\end{align}
Next, following Eq.~\eqref{fund} we can make manifest that $x_i$ and $x_f$ lie in different subsets of space by decomposing the propagator as
\begin{align}
    D_F & (x_i | x_f ,t ) = \hspace{-3pt}  \sint{0}{t}{0.2} {\rm d} t_s \ \bar{D}_F (x_i| x_s ,t_s ) D_F ( x_s , t_s  | x_f , T), \label{Ddecomp}
\end{align}
where $x_s$ is the closest point in $\Omega_{+}$ for which the particle's energy is degenerate with its energy at point $x_i$.\footnote{A similar decomposition is possible in the presence of more degrees of freedom, in which case one needs to include an additional integral of $x_s$ over a suitable hypersurface.} To define the auxiliary quantity $\bar{D}_F$, we may first introduce the functional $F_{x_s}[x]$, which (as noted above) maps any time-dependent path $x$ onto the time when it first reaches $x_s \subset \Omega_{+}$. In terms of this object, $\bar{D}_F$ is defined as
\begin{equation}
    \bar{D}_F (x_i | x_s , t_s ) \equiv \mathcal{N} \sint{x (0) = x_i}{x (t_s) = x_s}{1} {\mathcal D} x \ e^{i S[x]} \delta \left( F_{x_s}[x] - t_s \right),
\end{equation}
with the usual path integral normalization factor $\mathcal{N}$. Moving forward, we will refer to the condition $F_{x_s}[x] = t_s$ as the \textit{crossing condition}. The decomposition in Eq.~\eqref{Ddecomp} amounts to splitting the time evolution of the particle's motion from $x_i$ to $x_f$ into a piece connecting $x_i$ with an energetically degenerate point in $R$ and the time evolution thereafter. Using this representation, the probability $P_{R}$ can be expressed as
\begin{align}
    P_{\Omega_{+}}=& \int_0^T \hspace{-0.4em}\text{d}t_s  \int_0^T \hspace{-0.4em}\text{d}t_s^{\prime} \,  \bar{D}_F (x_i | x_s ,t_s ) \bar{D}_F^* (x_i | x_s , t_s^\prime ) \nonumber \\
    & \times \int_{\lowersub \Omega_+} \hspace{-0.4em}  {\rm d} x_f \, D_F(x_s, t_s | x_f, T)  D_F( x_f, T | x_s, t_s^\prime)  . \label{PRlong}
\end{align}
For the decay through a single tunneling event no back-tunneling occurs, which allows us simplify Eq.~\eqref{PRlong} as
\begin{align}
      \int_{\Omega_{+}}  {\rm d} x_f \, &D_F(x_s, t_s | x_f, T)  D_F( x_f, T  | x_s, t_s^\prime)   \nonumber \\ 
      &\simeq D_F( x_s, t_s | x_s, t_s^\prime) = D_F^*(x_s, t_s^\prime | x_s, t_s ) .\label{Dss}
\end{align}
Next, we can rewrite the time integrals as
\begin{equation}
    \int_0^T \text{d}s  \int_0^T \text{d}s^{\prime} = \int_0^T \text{d}s  \int_0^s \text{d}s^{\prime} + \int_0^T \text{d}s^{\prime} \int_0^{s^{\prime}} \text{d}s  .
\end{equation}
Doing so allows us to use the second integral in each of these combinations to recombine the propagator in Eq.~\eqref{Dss} with one of the two factors of $\bar{D}_F$ using Eq.~\eqref{Ddecomp}. Eliminating the remaining time integral by taking a derivative, we are ultimately left with 
\begin{align}
    \frac{{\rm d} P_{\Omega_{+}}}{ {\rm d} t}= \bar{D}_F(x_i| x_s, T) D_F^* (x_s, T | x_i) + c.c.
\end{align}
Hence we find that the decay rate can be represented through the simple path-integral expression
\begin{align}
    \Gamma =& \lim_{T \to \infty} \frac{|\mathcal{N}|^2}{Z (T)} \sint{x_1 (0)=x_i}{x_1 (T)=x_s}{0} {\mathcal D} x_1 \ e^{i S[x_1] } \delta \left( F_{x_s}[x_1] - T \right)  \nonumber \\ 
    &\quad\quad\quad\quad\quad\quad\times \left(\ \ \ \sint{x_2 (0)=x_i}{x_2 (T)=x_s}{1} {\mathcal D} x_2 \  e^{ iS[x_2]} \right)^* + c.c.  \label{GammaMaster} 
\end{align}
Similarly, the normalization factor $Z$ can be represented as a product of two path integrals:
\begin{align}
    Z (T)=&P_{\Omega_{-}} (T)=  \int_{\Omega_{-}}  {\rm d} x_v \langle x_v,T | x_i\rangle \langle x_i | x_v,T \rangle \label{NormalizationMaster} \\ 
    =&  \int_{\Omega_{-}}  {\rm d} x_v \ D_F( x_v ,T | x_i ) D_F(x_i| x_v,T  ) \nonumber \\ 
    =& |\mathcal{N}|^2 \int_{\Omega_{-}}  {\rm d} x_v \sint{z_1 (0)=x_i}{z_1 (T)=x_v}{1} {\mathcal D} z_1 \ e^{i S[z_1] }   \left(\ \ \ \sint{z_2 (0)=x_i}{z_2 (T)=x_v}{1} {\mathcal D} z_2 \  e^{ iS[z_2]} \right)^* . \nonumber
\end{align}
In the cases $x_i = x_{-}$ and $z_i = x_{-}$, these expressions can be simplified through Wick rotations, which translate the real-time actions to Euclidean ones. Crucially, this suggests that one identify the physical time $T$ with the duration of the imaginary-time interval $\tau$. Motivated by our previous discussion, we will instead evaluate these path integrals through a saddle-point approximation in real time.

In Secs.~\ref{sec:Steadyon} and ~\ref{sec:Drag} we show how to construct suitable stationary-phase solutions for the equations of motions, with the boundary conditions imposed by the path integral. However, even without having at hand the precise form of these solutions, it is easy to see that they are fully determined by their boundary conditions. Formally, this implies $\bar{x}_1=\bar{x}_2 \equiv \bar{x}$, where $\bar{x}_{1,2}$ is the stationary phase of the $x_{1,2}$ path integral. As the same holds true for the path integrals in the denominator involving paths $z_{1,2}$ with corresponding stationary phases $\bar{z}_{1,2}$ (restricted to paths that terminate within $\Omega_{-}$), we thus find that the decay rate is, to leading, of the form
\begin{align}
    \Gamma =&  A\cdot \frac{e^{iS[\bar{x}]-iS[\bar{x}]^* }}{e^{iS[\bar{z}]-iS[\bar{z}]^* }}= A\cdot e^{-\left( 2 {\rm Im} (S[\bar{x}])- 2 {\rm Im} (S[\bar{z}])\right)} .
    \label{eq:GammaUs}
\end{align}
Here $A$ denotes next-to-leading order corrections, which will not be addressed in this work. In other words, to leading order the tunneling rate is only sensitive to the imaginary part of the action evaluated on the stationary-phase solutions, which may take complex values due to the regularization of the path integral. In Secs.~\ref{sec:Steadyon} and~\ref{sec:Drag}, we show that, in the combined limit of a vanishing regulator and an infinite physical time, these quantities reproduce the familiar leading-order results for a system without an explicit time dependence,
\begin{align}
    \lim_{T \to \infty}\Gamma (T) = A\cdot \frac{e^{-S_E [\bar{x}_{ I}]}}{e^{-S_E [\bar{z}_{ I}]}} .
\end{align}
Here $\bar{x}_{ I}$ denotes the well-known periodic instanton defined on some Euclidean time interval $\Delta \tau_I$, while $\bar{z}_{ I}$ plays a similar role for the normalization factor.

\section{Regularized real-time dynamics}\label{sec:RegularizedDynamics}

In order to clarify the relation between real- and imaginary-time properties of the instanton mediating the decay, we would like to evaluate the path integrals in Eqs.~\eqref{GammaMaster} and \eqref{NormalizationMaster} directly. To avoid the conceptual difficulties linked to a Wick rotation, we will regularize the path integral while keeping the time explicitly real. This can be achieved by introducing a small imaginary part to the Hamiltonian, which amounts to the replacement
\begin{equation}\label{eq:Reg}
    H(t) \to (1 - i \epsilon) H(t).
\end{equation}
For the case in which $H$ is independent of time, the replacement of Eq.~(\ref{eq:Reg}) is equivalent to performing a Wick rotation. On the other hand, we may perform the substitution of Eq.~(\ref{eq:Reg}) even for systems with explicit time dependence, for which a simple Wick rotation is even more involved~\cite{Kaya:2018jdo}.

In terms of the point particle's action, Eq.~\eqref{eq:Reg} amounts to the replacement
\begin{align}\label{eq:actionreg}
    S\to (1 + i \epsilon) \intd t \ \frac{m}{2}\dot{x}^2 -  (1 - i \epsilon)^2 \cdot  V (x),
\end{align}
where, in this section, an overdot denotes a derivative with respect to $t$.

For concreteness, we may consider a point particle in a double-well potential, whose action we will take to be
\begin{align}
    S=& (1 + i \epsilon)  \int_0^t {\rm d} t^{\prime} \ \frac{m}{2} \dot{x}^2 - (1 - i \epsilon)^2 \cdot \frac{\lambda}{4}\left( x^2 - x_0^2 \right)^2   \nonumber\\ 
    =& x_0^3 \sqrt{\lambda m}  (1 + i \epsilon) \intd \tilde{t}^{\prime} \ \frac{1}{2}\dot{\tilde{x}}^2 - \frac{1}{4} (1 - i \epsilon)^2  \left( \tilde{x}^2 -1\right)^2  \nonumber \\ 
    =& x_0^3 \sqrt{\lambda m} \cdot \tilde{S} . \label{SDW}
\end{align}
In order to obtain the action in the second line, we have rescaled the time and position variables as $x=\tilde{x} \cdot x_0$ and $t = \sqrt{m/\lambda} \cdot \tilde{t}/x_0$. Moving forward, we will drop the tilde on all variables for simplicity.

Using Eq.~\eqref{SDW}, it is straightforward to arrive at the equation of motion for the (complex-valued) saddle point,
\begin{equation}
    \ddot{x}=- (1-2 i \epsilon) x \left(x^2 -1\right) .\label{compeom}
\end{equation}
This equation once again confirms the equivalence with the interpretation of the imaginary part as a result of a complex-time contour, as the factor $(1-2 i \epsilon)$ on the right-hand side could equivalently be moved to the left-hand side of the equation and absorbed into the time derivatives through an infinitesimal Wick rotation $t^{\prime} \to (1- i \epsilon) \cdot t^{\prime}$.

The solutions of Eq.~\eqref{compeom} that dominate the path integrals in Eqs.~\eqref{GammaMaster}--\eqref{NormalizationMaster} are now determined by their respective boundary conditions,
\begin{gather}  
    \bar{x}_{1,2}(0)=x_i, \ \  \ \bar{x}_{1,2}(T)=x_s ,   \label{eq:xbc} \\
        \bar{z}_{1,2}(0)=x_i, \ \  \ \bar{z}_{1,2}(T)=x_v  ,\label{eq:zbc}
\end{gather}
as well as the crossing condition for $\bar{x}_1$: $F_{x_s}[\bar{x}_1]=T$.

Eq.~\eqref{compeom} has well-known analytic solutions, which can be expressed in terms of the Jacobian elliptic functions ${\rm sn} (u, k^2)$ as
\begin{equation}\label{sol}
    x_k (t)= \sqrt{\frac{2 k^2}{1+k^2}} \, {\rm sn}\left( \frac{i e^{-i \epsilon} \cdot t - b}{\sqrt{1+k^2}}, k^2 \right).
\end{equation}
The parameter $k$, known as the elliptic modulus, controls both the solution's amplitude and oscillation frequency, while the parameter $b$ corresponds to a translation.

For $\epsilon \to 0$, these solutions reproduce the behavior of a point particle in the physical potential. In the limit $\epsilon \to \pi/2$, meanwhile, they describe the motion of a particle in the inverted potential, including, in particular, the periodic instantons obtained in the conventional picture~\cite{Liang:1992ms,Liang:1994xn,Liang:1994qr,Liang:1998hk,Muller-Kirsten:2001jhw,Muller-Kirsten:2012wla}.

A similarly simple interpretation can be obtained for the case of a small but finite $\epsilon$, for which the path can be decomposed into a real and imaginary part, $\bar{x}(t)= \bar{x}_{\rm Re} (t) + i \bar{x}_{\rm Im} (t)$. Their corresponding equations of motions are given by
\begin{align*}
    \ddot{\bar{x}}_{\rm Re}=& \bar{x}_{\rm Re} \left(1-\bar{x}_{\rm Re}^2  + 3 \bar{x}_{\rm Im}^2 \right) + 2 \epsilon \cdot \bar{x}_{\rm Im} \left( 1+ \bar{x}_{\rm Im}^2 - 3 \bar{x}_{\rm Re}^2 \right), \\
    \ddot{\bar{x}}_{\rm Im}=& \bar{x}_{\rm Im} \left(1+\bar{x}_{\rm Im}^2 - 3 \bar{x}_{\rm Re}^2 \right) + 2 \epsilon \cdot \bar{x}_{\rm Re} \left( \bar{x}_{\rm Re}^2 -1 -3 \bar{x}_{\rm Im}^2 \right).
\end{align*}
For small $t$ relative to the system's typical time-scale, when $\bar{x}_{\rm Im} (t) \simeq 0$, the system acts essentially classically, leading to an oscillatory motion of $\bar{x}_{\rm Re} (t)$ in the false-vacuum basin. This motion sources an excitation of the imaginary part $\bar{x}_{\rm Im} (t)$ through the interactions conveyed by a non-vanishing $\epsilon$, which in turn backreacts on $\bar{x}_{\rm Re} (t)$. This process can unfold in two physically relevant ways. In Sec.~\ref{sec:Steadyon}, we construct the real-time equivalent of the well-known instanton satisfying the boundary conditions of Eq.~\eqref{eq:xbc}, which we refer to as \textit{steadyons}. Then, in Sec.~\ref{sec:Drag}, we discuss solutions satisfying the boundary conditions in Eq.~\eqref{eq:zbc}, which are imposed by the normalization factor $Z$.

Lastly, we note that, as for all path integrals, the stationary-phase method we employ is only capable of capturing the contribution arising from the neighborhood of one particular configuration, or at best a finite-dimensional submanifold of path space in theories giving rise to collective coordinates. This aspect becomes even more subtle in our picture. If one did not regularize the path integral, there would not exist {\it any} stationary-phase solutions. For this reason, the regularization of Eq.~(\ref{eq:Reg}) can be understood as deforming the theory in such a way as to ``create'' a viable stationary-phase solution at one particular point in path space, which in turn allows for analytical access to its neighborhood.\footnote{Something similar is already necessary to calculate the decay rate of the electroweak vacuum in the Standard Model. As discussed in Ref.~\cite{Andreassen:2017rzq}, no instanton exists for the full potential of this theory due to the Higgs field's mass term, despite its small numerical value relative to the relevant scales of the process. This problem can, however, be resolved through a deformation of the theory by removing the problematic mass term entirely. It can then be shown that this does, indeed, induce an error to the total decay rate, which vanishes in the limit $m^2 \to 0$.}

In Sec.~\ref{sec:NPSteadyonSolution}, we show that our procedure can be used to construct stationary phases along a one-parameter family of points in path space. In the imaginary-time picture, these points correspond to different initial momenta of their corresponding instantons.

\section{Steadyon contribution}\label{sec:Steadyon}

The real-time equivalent of an instanton would correspond to a solution to the equations of motion of the form in Eq.~\eqref{sol}, together with the boundary conditions in Eq.~\eqref{eq:xbc}. Crucially, these conditions do not specify the initial velocities of $\bar{x}_{\rm Re} (t)$ and $\bar{x}_{\rm Im} (t)$, much as the initial velocity of an instanton solution is undetermined in the usual imaginary-time picture. For tunneling out of the false vacuum, this degeneracy is eliminated through the infinite length of the imaginary-time interval, as laid out in great detail in Ref.~\cite{Andreassen:2016cvx}. It is now easy to see that, even if one were to start out in the imaginary-time picture used in Ref.~\cite{Andreassen:2016cvx}, the same procedure cannot be applied to the case of interest in this article. 

We therefore consider two different classes of solutions for the real-time dynamics, which we dub \textit{steadyons}. We adopt this term because such real-time solutions describe a continuous transition from the initial state to the desired final state, over an infinitely long physical time. We first consider periodic steadyon solutions in Sec.~\ref{sec:SteadyonSolution}, corresponding to vanishing initial velocity, before generalizing to the case of nonvanishing initial velocities in Sec.~\ref{sec:NPSteadyonSolution}. As we will see, in the latter case, such solutions include the dominant contribution to the decay rate.

\subsection{Periodic steadyon solutions}\label{sec:SteadyonSolution}

\begin{figure*}[t!]
    \centering
    \includegraphics[width=0.32\textwidth]{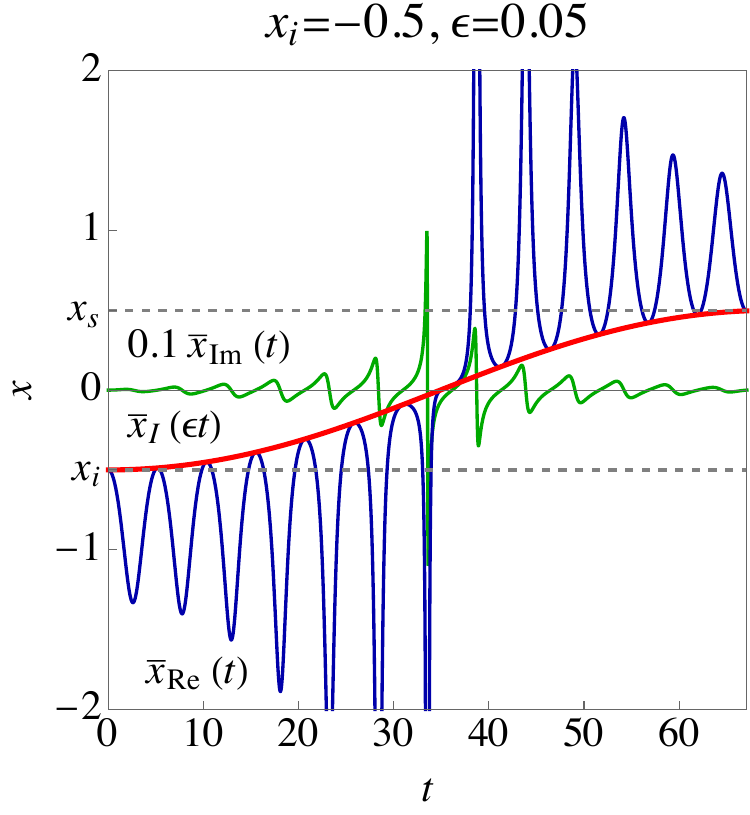}
    \includegraphics[width=0.32\textwidth]{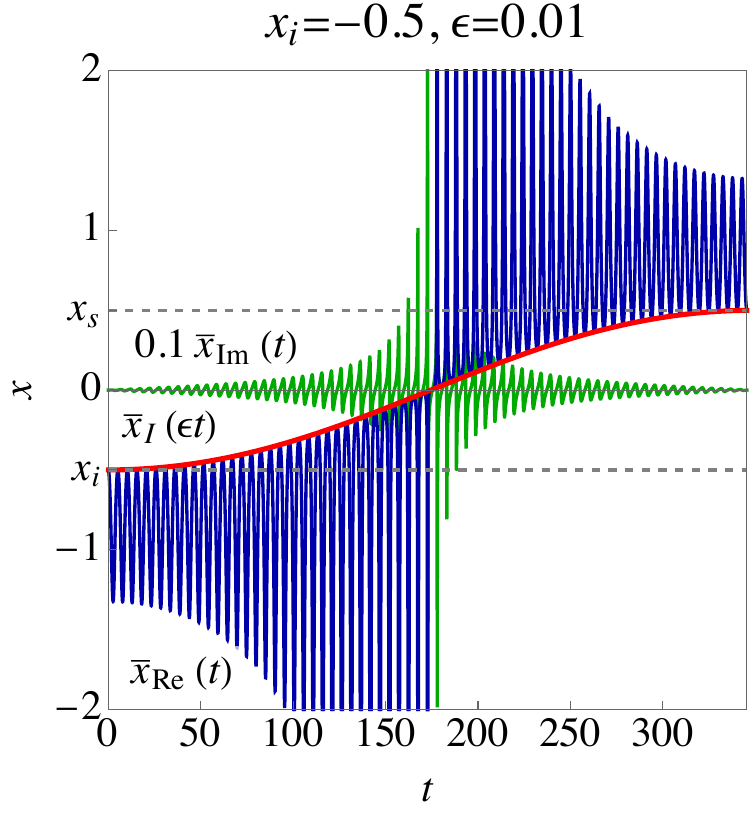}
    \includegraphics[width=0.32\textwidth]{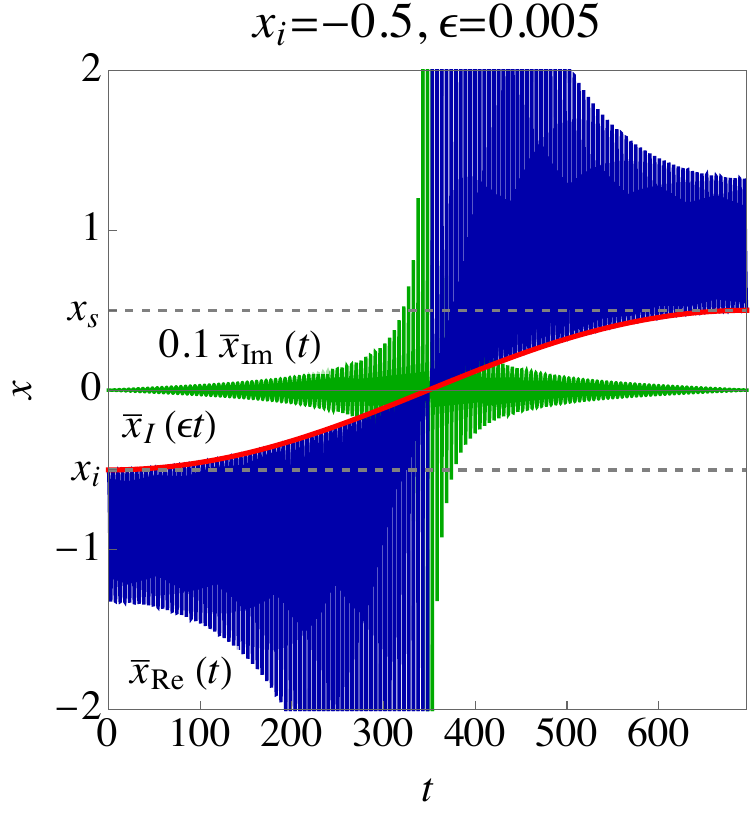}
    \caption{The periodic steadyon solution for $x_i=-1/2$ and $b$ given by Eq.~\eqref{tau0ofk}.}
    \label{fig:ESCI}
\end{figure*}

In Sec.~\ref{sec:ComplexTime}, we found that the real- and imaginary-time versions of the instanton describing tunneling out of the false vacuum in a system without explicit time-dependence were related to one another through a simple analytic continuation. For now focusing on the special case of a periodic solution, this suggests the choice of parameters
\begin{gather}
    k^2 = \frac{x_i^2}{2-x_i^2} , \label{k2ofxi}\\ 
    b =  \sqrt{1 + k^2} \int_0^{\frac{\pi}{2}}{\rm d} \theta \frac{1}{\sqrt{1-k^2 \sin (\theta^2)}}. \label{tau0ofk}
\end{gather}
An example of such solutions is shown in Fig.~\ref{fig:ESCI}. 

The solutions in Fig.~\ref{fig:ESCI} can be understood easily in terms of our discussion in the previous section. At $t=0$, the particle sets out at the appropriate initial position with a vanishing velocity for both $\bar{x}_{\rm Re}$ and $\bar{x}_{\rm Im}$. The oscillatory motion of $\bar{x}_{\rm Re}$ then causes an excitation of $\bar{x}_{\rm Im}$, whose backreaction on $\bar{x}_{\rm Re}$ ultimately causes the latter to leave the false-vacuum basin, after which the backreaction starts damping the oscillation of $\bar{x}_{\rm Re}$. Eventually, $\bar{x}_{\rm Im}$ once again vanishes, as $\bar{x}_{\rm Re}$ transitions back to its classical periodic motion, albeit now within the true-vacuum basin. Physically, this suggests that the transition from one basin to the other is controlled by the product $\epsilon \cdot t$. Understanding the regularization as the introduction of a complex time variable, this combination can be identified with the change in imaginary time along the complex-time contour, in agreement with our observation regarding the tunneling out of the minimum in Sec.~\ref{sec:ComplexTime}.

Indeed, when understood as analytic functions of a complex argument $i e^{- i \epsilon} t$, the solutions of Eq.~\eqref{sol} are doubly-periodic in the complex plane. This relates, in particular, to the period of the periodic instanton $\Delta \tau_I$, which can immediately be identified with the period of the solution in Eq.~\eqref{sol} along the imaginary-time axis. Meanwhile, the period along the real-time axis can be recognized as the period of the oscillatory motion within the false-vacuum basin in real time. Moving forward, we will denote the period of this classical motion as $\Delta t_R$. 

The transition of the particle from one basin to the other amounts to an evolution through one-half period of the periodic motion along the imaginary-time direction. Together with the crossing condition, this implies that a transition from $x_i$ to $x_s$ after the time $T$ would require
\begin{gather}
    \epsilon \cdot T = \frac{\Delta \tau_I}{2} \ \ {\rm and}  \label{Asynch1} \\ 
     T = N \cdot \Delta t_R, \ \ {\rm with} \ \ N \in \mathbb{N}.  \label{Asynch2}
\end{gather}
Here $\Delta \tau_I$ denotes the full period of the periodic (imaginary-time) instanton connecting $x_i$ to $x_s$, while $\Delta t_R$ is the duration of the particle's classical oscillatory motion within the false-vacuum basin. Note that, in general, both of these parameters depend on $x_i$. 

\begin{figure*}[t!]
    \centering
    \includegraphics[width=0.32\textwidth]{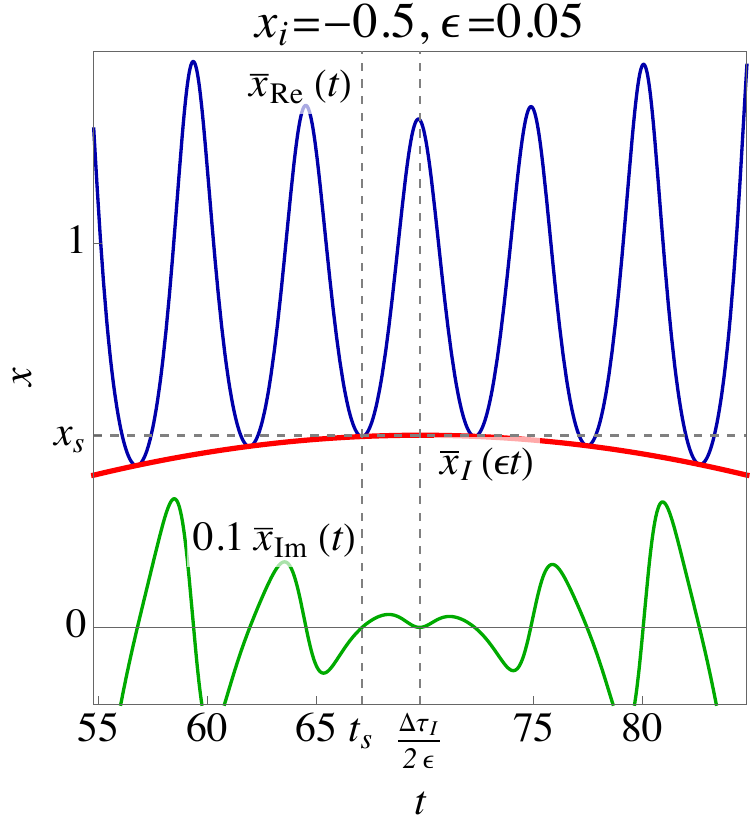}
    \includegraphics[width=0.32\textwidth]{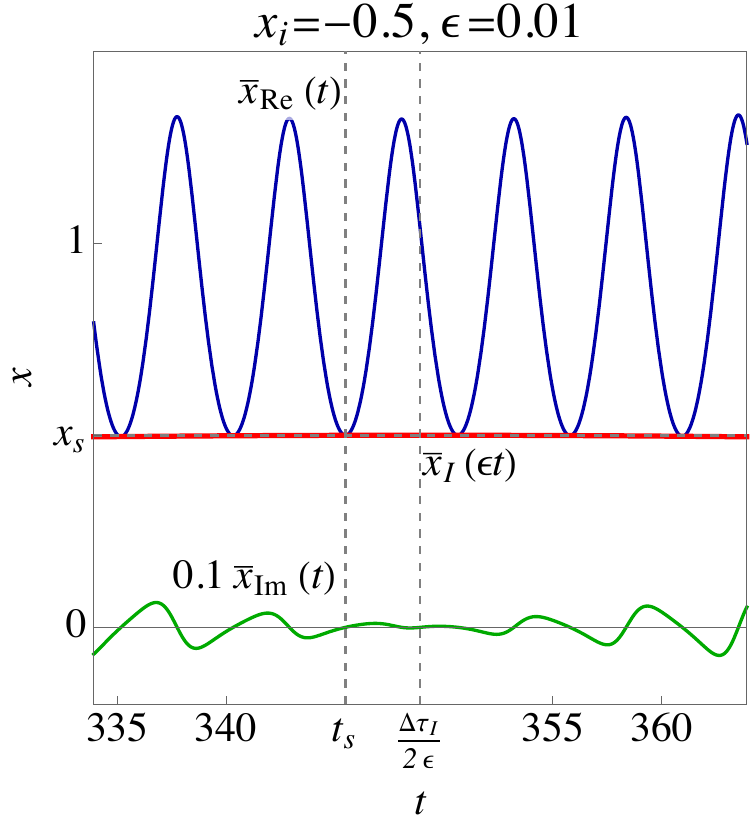}
    \includegraphics[width=0.32\textwidth]{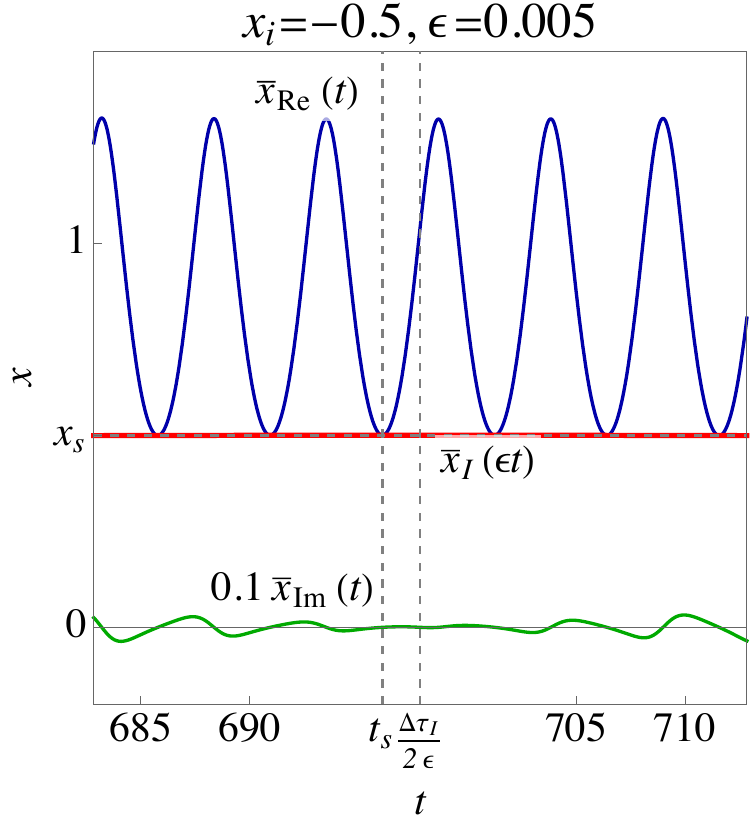}
    \caption{The periodic steadyon around $\Delta \tau_I/(2 \epsilon)$ for $x_i=-1/2$ and different values of $\epsilon$. For $t_s = N \cdot \Delta t_R$ with $N \in \mathbb{N}$, $\bar{x}_{\rm Re}$ coincides with the value of the corresponding instanton $\bar{x}_I$ evaluated at $\tau = \epsilon \cdot t_s$, and $\bar{x}_{\rm Im}$ vanishes. These times, however, do \textit{not} necessarily agree with the physical time $T$, as would be required by the boundary conditions in Eqs.~(\ref{eq:xbc})--(\ref{eq:zbc}). Thus, for any given $\epsilon$, we may restrict ourselves to $T=t_s$. In that case, the boundary condition is not satisfied exactly, but only up to some offset~$\propto \epsilon$, which vanishes in the limits of interest.}
    \label{fig:ESBC}
\end{figure*}

These conditions are clearly not satisfied simultaneously for arbitrary times $T$. In other words, for general $T$, the solutions described by Eq.~\eqref{sol} fail to reproduce the correct boundary conditions of Eqs.~\eqref{eq:xbc}--\eqref{eq:zbc}, as illustrated in Fig.~\ref{fig:ESBC}. This behavior can be easily understood through the splitting of the dynamics into an independent real and imaginary part. There is {\it a priori} no reason for the oscillatory motion of the real part to arrive at the location $x_s$ at the very moment when the imaginary part vanishes identically. Even worse, as the period of the real part's dynamics is independent of $\epsilon$, Eq.~\eqref{Asynch2} seems to suggest that the steadyon can only satisfy the appropriate boundary condition during a countable subset of physical times.\footnote{This behavior is a consequence of our choice of the initial state together with the semi-classical picture, which assumes that the tunneling occurs out of the classical motion's turning point.}

This is, however, not surprising, but rather a manifestation of the physical motivation for our ansatz in Eq.~\eqref{Gammadef}. Recall that we had argued that the loss of probability within $\Omega_-$ should happen predominantly at times for which the point particle is furthest up the potential barrier, i.e., $T= N \cdot \Delta t_R$. And indeed, this is precisely the behavior shown by the steadyon.

\begin{figure}[h!]
    \centering
    \includegraphics[width=0.48\textwidth]{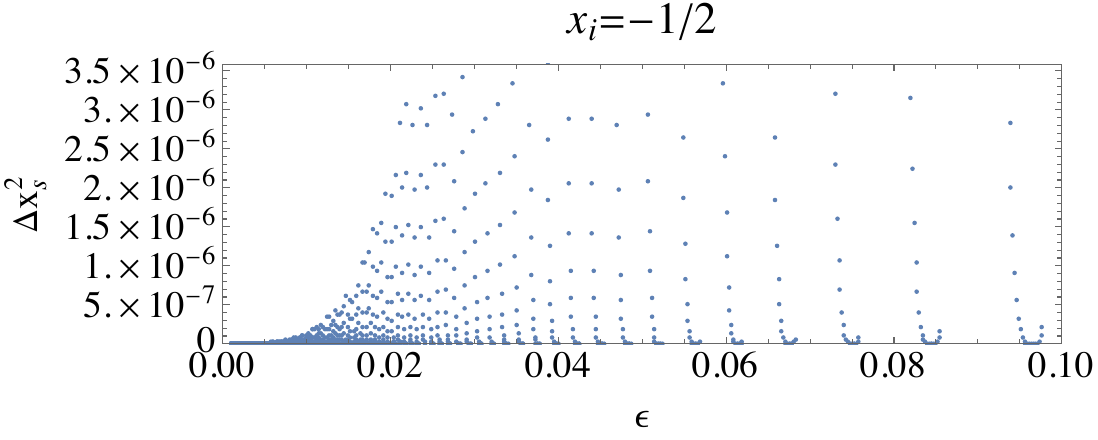}
    \caption{Convergence of the boundary condition at $t_s$ in the limit $\epsilon \to 0$ for the periodic steadyon.}
    \label{fig:dxs}
\end{figure}

Following our previous discussion, the emergence of a non-vanishing imaginary part $\bar{x}_{\rm Im} (t)$ can be understood as induced through its interaction with the real part $\bar{x}_{\rm Re} (t)$, whose strength is controlled by $\epsilon$. As an immediate consequence, taking the limit $\epsilon \to 0$ implies that the amplitude of $\bar{x}_{\rm Im} (t)$ vanishes in the vicinity of its root, as illustrated in Fig.~\ref{fig:ESBC}. In other words, in the limit $\epsilon \to 0$ the imaginary part $\bar{x}_{\rm Im} (t)$ vanishes within a time interval roughly of length $\Delta t_R$. Thus, the boundary conditions of Eqs.~\eqref{eq:xbc} and~\eqref{eq:zbc} can be approximately satisfied by identifying $T$ with the time $t_s$ closest to $\Delta \tau_I / (2 \epsilon)$ at which $\bar{x} (t_s) \simeq x_s$ with the highest accuracy. A sensible measure to define the deviation of the boundary conditions is $\Delta x_s^2 = (\bar{x}_{\rm Re}(T)-x_s)^2 + \bar{x}_{\rm Im}(T)^2$.\footnote{This choice, of course, is arbitrary, and in Sec.~\ref{sec:SteadyonAction} we demonstrate that our result for the tunneling rate is independent of how we choose to quantify the departure from exact boundary conditions, as long as $\Delta x_s^2 \to 0$ in the appropriate limit. In cases in which there exist multiple values of $t_s$ that yield a similar convergence rate, we choose the smallest such value, as suggested by the crossing condition.} And indeed, as shown in Fig.~\ref{fig:dxs} for the concrete example corresponding to $x_i=-1/2$, taking the limit $\epsilon \to 0$ and choosing $T=t_s$ leads to $\Delta x_s \to 0$.
\begin{figure}[h!]
    \centering
    \includegraphics[width=0.48\textwidth]{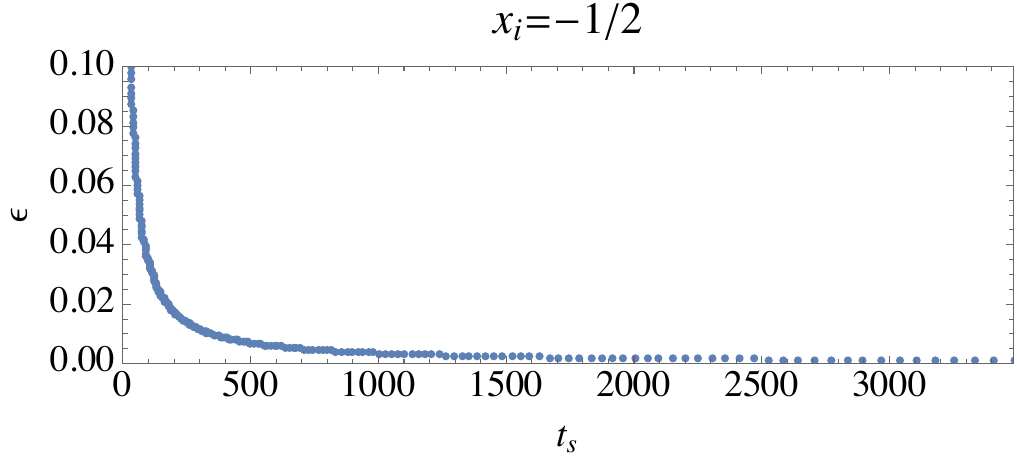}
    \caption{The relation between the time $t_s$ and the regularization parameter $\epsilon$ for the periodic steadyon.}
    \label{fig:tse}
\end{figure}

This behavior can equivalently be understood by observing that $\bar{x}_{\rm Re}(t)=\bar{x}_{ I}(\epsilon \cdot t)$ whenever $\bar{x}_{\rm Im} (t) =0$. In the limit $\epsilon \to 0$, the time $t_s$ coincides with the time closest to $\Delta \tau_I/(2 \epsilon)$ where this occurs. However, since the evolution of $\bar{x}_{I} (t)$ is controlled by the combination $\epsilon \cdot t$, it is easy to see that the residual difference between $\bar{x}_{\rm Re}(t_s)$ and $x_s$ satisfies
\begin{align}
    \bar{x}_{\rm Re}(t_s)=&\, \bar{x}_{I}(\epsilon \cdot t_s)   \label{eq:dxexp} \\ 
    =&\, x_s -|\dot{\bar{x}}_{ I}(\Delta \tau_I/2) |  \cdot \epsilon \cdot  | t_s - \Delta \tau_I/(2\epsilon )|   \nonumber\\
    &-\frac{1}{2} |\ddot{\bar{x}}_{ I}(\Delta \tau_I/2) | \cdot \epsilon^2 \cdot  | t_s - \Delta \tau_I/(2\epsilon )|^2 + \mathcal{O}(\epsilon^3). \nonumber 
\end{align}
By construction $| t_s - \Delta \tau_I/(2\epsilon )| \leq \Delta t_R = {\rm constant}$, so we find that the boundary value of the steadyon at $t=t_s$ indeed converges toward the desired emergence point $x_s$. For the periodic steadyons, this convergence is further accelerated, given that $\dot{\bar{x}}_{I}(\Delta \tau_I/2) =0$.

We may now recall that one of the main motivations for our analysis was the simple observation that the usual interpretation of imaginary time as Wick-rotated physical time seemed to suggest that an instanton can only exist at one finite physical time, of order of the system's natural time-scale. At first glance, this appears to be true even in our more careful treatment, as there is still only one unique value for $t_s$ for any given $\epsilon$. This apparent problem can be resolved thanks to two simple observations. First, we note that the relation between $x_i$, $t_s$, and $\epsilon$ necessary to (approximately) satisfy the boundary condition can be understood equivalently as determining $\epsilon$ in terms of the physical parameters $x_i$ and $t_s$, with $\epsilon \sim \Delta \tau_I/(2 t_s)$. Second, we recall that in our picture, $\epsilon$ is not an \textit{entirely} free parameter, but corresponds to a regularization of the Hamiltonian, and hence needs to be taken to $0$ at the end of a given calculation. These two points suggest that for any suitable time $t_s$ of interest, we may choose the ``right'' regularization parameter $\epsilon$ to ensure the existence of an approximate stationary phase. Inevitably, any such choice implies a finite error. However, as we consider larger and larger physical times (in discrete steps), we also lower the value of $\epsilon$ required, implying an increasingly better approximation. In the limit $T=t_s \to \infty$, this automatically yields $\epsilon \to 0$. See Fig.~\ref{fig:tse}.

\subsection{Non-periodic steadyon solutions}\label{sec:NPSteadyonSolution}

\begin{figure*}[t!]
    \centering
    \includegraphics[width=0.32\textwidth]{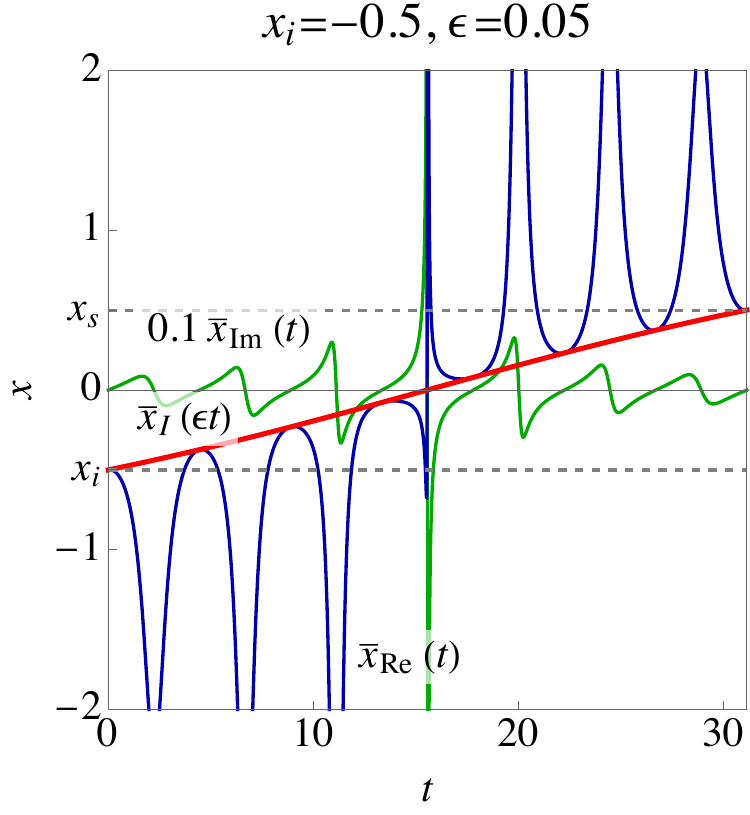}
    \includegraphics[width=0.32\textwidth]{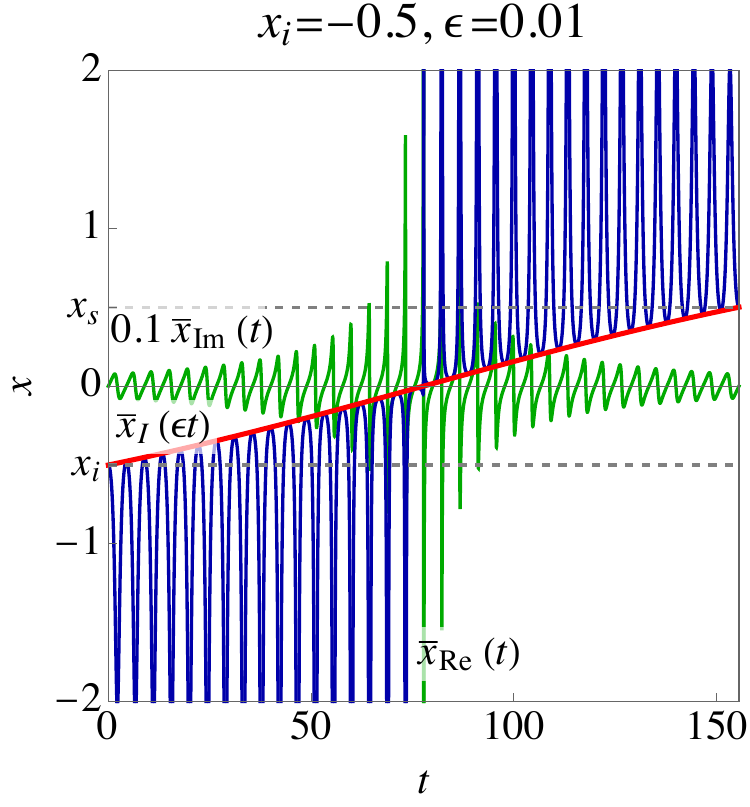}
    \includegraphics[width=0.32\textwidth]{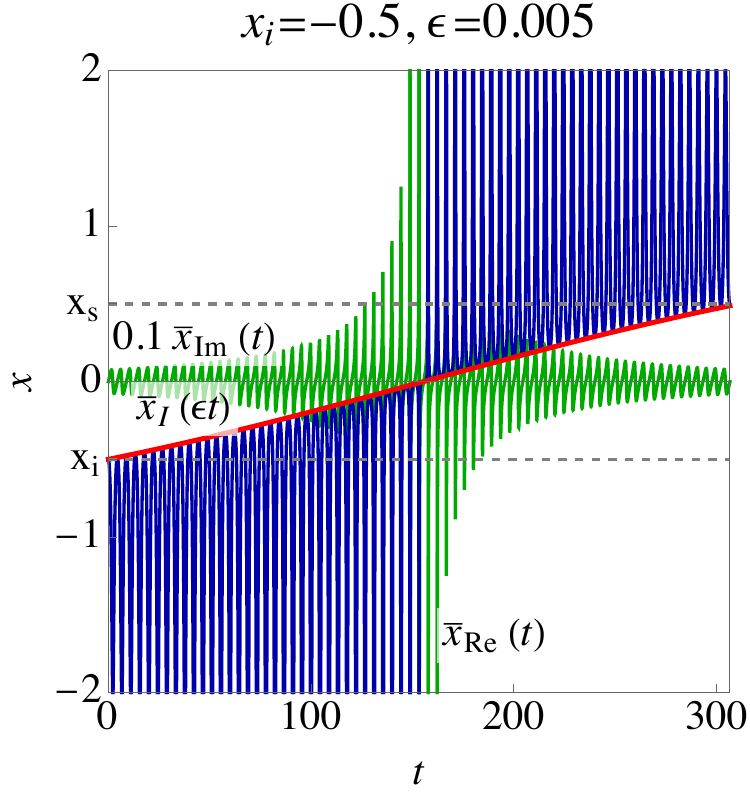}
    \caption{The steadyon describing the same tunneling process as in Fig.~\ref{fig:ESCI}, but with the initial velocity $\dot{\bar{x}}_{\rm Im} (0) \neq 0$ corresponding to the largest contribution to the decay rate.}
    \label{fig:EStCI}
\end{figure*}

Having understood the special case of a periodic steadyon solution, we can now generalize our previous discussion to cases with an initial velocity. An important feature of the solutions of Eq.~\eqref{sol} is that their imaginary part vanishes exactly at the turning points of the real part, as shown, e.g., in Fig.~\ref{fig:ESBC}. Therefore the boundary conditions of Eq.~\eqref{eq:xbc} require $\dot{\bar{x}}_{\rm Re}(0)=\dot{\bar{x}}_{\rm Re}(T)$, leaving only $\dot{\bar{x}}_{\rm Im}(0)$ as a free parameter. 

Returning to our interpretation of the steadyon as two interacting degrees of freedom, we can immediately understand the effect of this initial velocity. For the periodic solution, we had found that the tunneling occurs as the oscillations of $\bar{x}_{\rm Re} (t)$ and $\bar{x}_{\rm Im} (t)$ drive each other until $\bar{x}_{\rm Re} (t)$ crosses from the false-vacuum basin $\Omega_{-}$ into the other well $\Omega_{+}$. Upon giving $\bar{x}_{\rm Im} (t)$ an initial velocity, this process will unfold faster for a given value of $\epsilon$, also leading to a smaller $\Delta \tau_I$. See Fig.~\ref{fig:EStCI}. 

Again making use of the connection between steadyon solutions and their corresponding instantons, this suggests that the steadyon with finite initial velocity $\dot{\bar{x}}_{\rm Im}(0)$ can also be described by a solution of the form in Eq.~\eqref{sol}, albeit with a larger value of $k$. In Sec.~\ref{sec:SteadyonAction}, we will find that for any $x_i$ the largest contribution to the decay rate is given by $k=1$, which corresponds to the false-vacuum instanton (or, equivalently, steadyon). In other words, the solution leading to the most important contribution to the overall decay rate is \textit{the same} as for the tunneling out of the false vacuum, defined on a shorter interval $\Delta \tau_I$ to match the desired boundary conditions of Eq.~\eqref{eq:xbc}. This is precisely the case illustrated in Fig.~\ref{fig:EStCI}.

The realization of these boundary conditions can be achieved in a similar way as for the periodic case. For general physical times, no exact solution exists, in agreement with our initial ansatz. Changing the initial momentum of the instanton also changes the length of the interval $\Delta t_R$. This manifests in our calculation through the condition that the steadyon only exists for precisely these times. We can also understand this behavior more generally in terms of the usual interpretation of the path integral as a superposition of different paths. The position eigenstate has no definite momentum, such that the paths within the path integral setting out from that position contain all possible momenta. Thus, the simple picture given in Fig.~\ref{fig:Potential} can be understood as describing just one of these paths, for which the particle's momentum at $x=x_i$ vanishes. 

\subsection{Real-time action and instanton limit}\label{sec:SteadyonAction}

In the previous subsection, we have found that the steadyon only serves as a stationary phase in the limit $\epsilon \to 0$, and even then only for a countable number of points in time. First, we observe that the non-existence of a stationary phase \textit{only} implies that we cannot use the stationary-phase approximation to evaluate the path integrals at hand, whereas the decay rate itself nonetheless has a well-defined value, which could be calculated, e.g., on a lattice. Next, we are primarily interested in the rate in the combined limit $\epsilon \to 0$, $T \to \infty$. For this purpose, being only able to calculate the decay rate for a countable subset of times is sufficient as long as our results allow us to identify an unambiguous asymptotic value. 

\begin{figure}[h!]
    \centering
    \includegraphics[width=0.48\textwidth]{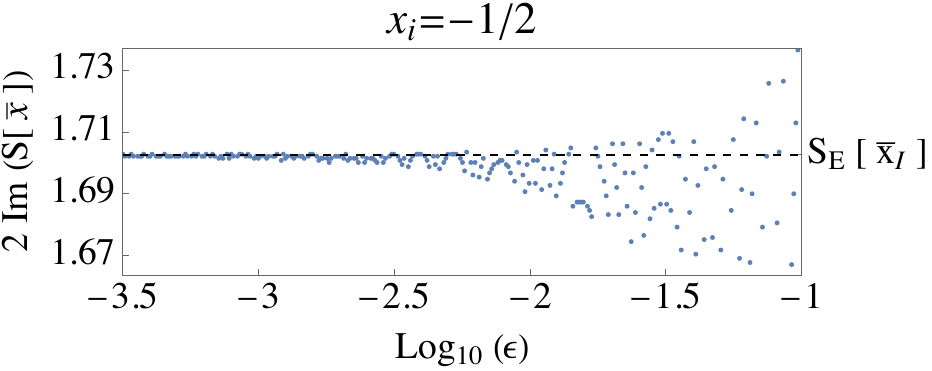}
    \caption{The exponent of the decay rate arising from the periodic steadyon contribution. We recover the Euclidean result in the appropriate limits.}
    \label{fig:SRSE}
    \includegraphics[width=0.48\textwidth]{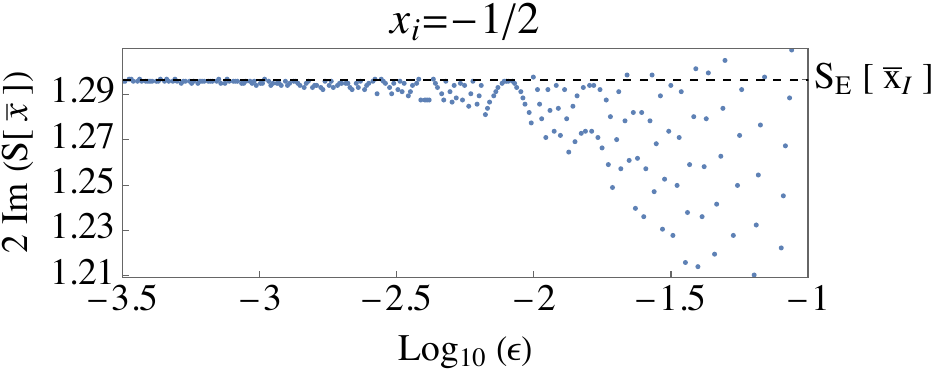}
    \caption{The exponent of the decay rate arising from the steadyon contribution with nonzero initial velocity corresponding to the largest contribution to the decay rate. We again recover the Euclidean result in the appropriate limits.}
    \label{fig:SRSEt}
\end{figure}

In order to illustrate this behavior, we calculate the imaginary part of the steadyon action for multiple values of $\epsilon$ by simply inserting the corresponding solutions into the regularized action of Eq.~\eqref{eq:actionreg} with the appropriate $t_s (\epsilon)$ for both the periodic steadyon as well as the steadyon with nonzero initial velocity, which we will later find to dominate the tunneling rate. And indeed, taking the limit in the suggested way leads to a clear convergence. Crucially, we find that the limiting value is given precisely by the Euclidean action of the corresponding instanton in the imaginary-time picture. See Figs.~\ref{fig:SRSE} and \ref{fig:SRSEt}.

\begin{figure}[t!]
\begin{tikzpicture}[scale=2.2]
    \draw[->,gray!60, ultra thick,dashed](2.5,-0.05) -- (2.5,-0.35);
    \draw[->,gray!60, ultra thick,dashed](2.5,-0.35) -- (2.5,-0.65);
    \draw[->,gray!60, ultra thick,dashed](2.5,-0.65) -- (2.5,-1.2);   
    \draw[->](0,0)--(2.7,0); 
    \draw[<-](0,-1.35)--(0,0);
    \draw[->, gray!60, ultra thick,dashed](0.05,0)--(2.5,0);
    \draw[->, gray!60, ultra thick,dashed](0.05,0)--(0.5,0);
    \draw[->, gray!60, ultra thick,dashed](0.05,0)--(1,0);
    \draw[->, gray!60, ultra thick,dashed](0.05,0)--(1.5,0);
    \draw[->, gray!60, ultra thick,dashed](0.05,0)--(2,0);
    \draw[black, thick](-0.2,-1.2)--(-0.2,0);
    \draw[black, dashed](-0.3,0)--(0,0);
    \draw[black, thick](2.8,-1.2)--(2.8,0);
    \draw[black, dashed](2.5,0)--(2.8,0);
    \draw[black, dashed](-0.3,-1.2)--(2.8,-1.2);
    \draw[black, dashed](-0.3,-1)--(2.5,-1);
    \draw[black, thick](2.5,-0.03)--(2.5,0.03);
    \draw[gray!60,  ultra thick](2.5,-1)--(0.05,-0.05);
    \draw[->, gray!60,  ultra thick](0.05,-0.05)--(1.5,-0.61);
    \draw[->, gray!60,  ultra thick](0.05,-0.05)--(1,-0.417);
    \draw[->, gray!60,  ultra thick](0.05,-0.05)--(2,-0.8);
    \draw (2.5,0.12) node [black]{$t_s$};
    \draw (2.7,0.15) node [black]
                       {$t$};
    \draw (-0.1,-1.3) node [black]
                       {$\tau$};
    \draw (1.45,0.15) node [black]
                       {Im$\left(S\right)=0$};
    \draw (-0.5,-0.5) node [black]
                       {$t_s \cdot \epsilon$};
    \draw (-0.5,-1.1) node [black]
                       {$O(\epsilon)$};
    \draw (3.1,-0.6) node [black]{$\Delta \tau_I/2$};
    \draw (0.35,-0.075) node [black]
                       {$\epsilon$};
    \draw [gray!60,ultra thick](0.5,0) arc [start angle=0, end angle=-60, radius=0.2];
\end{tikzpicture}
\caption{The exponent of the decay rate arises from the integral of the steadyon's complex Lagrange function over the diagonal time contour $\gamma_\epsilon$. This contour can be deformed into pieces along the real- and imaginary-time axis, respectively. The length of the imaginary-time piece coincides with $\Delta \tau_I/2$ up to a factor of order of the regularization parameter $\epsilon$.}
\label{fig:DeltaTauOrigin}
\end{figure}
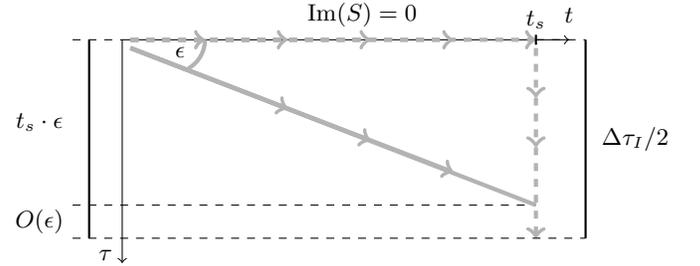

To understand this behavior, we reconsider the regularized action of Eq.~\eqref{eq:actionreg}, the resulting equation of motion in Eq.~\eqref{compeom}, and its analytic solutions in Eq.~\eqref{sol}. We note that, due to the time-independence of the Hamiltonian in this case, we could \textit{formally} assign the small imaginary part to the time variable, which can be understood as evaluating all functions along the complex-time path $\gamma_\epsilon (t)=(1 -i \epsilon) t$ rather than along the real-time axis. This suggests that we may understand the steadyon solutions, as well as their Lagrange functions, as their analytic continuations evaluated along this contour. Most important, these functions are regular and analytic between $\gamma_\epsilon$ and the real- and imaginary-time axes. Thus, we may deform the integration contour from $\gamma_\epsilon$ onto these axes, as shown in Fig.~\ref{fig:DeltaTauOrigin}. 

We first consider the contribution from the real axis, along which the steadyon reduces to the solution of the \textit{classical} equation of motion. This implies that its corresponding action is strictly real. Thus, this part of the contour does \textit{not} contribute to the leading-order exponent of the tunneling rate, which, as we found in Eq.~\eqref{eq:GammaUs}, depends only on ${\rm Im} (S [ \bar{x}])$.

Along the imaginary-time axis, the steadyon reduces to the familiar instanton solution, with an additional factor of $i$ coming from the integration along the imaginary-time direction. Following our previous discussion, the imaginary-time integral covers the interval $[0,t_s (\epsilon) \cdot \epsilon]$. The double-periodicity of the Jacobian elliptic functions then implies that the initial value of the instanton at $\tau=0$ is given by $x_i$ in the limit $\epsilon \to 0$. Meanwhile, by construction we have $|t_s (\epsilon) \cdot \epsilon - \Delta \tau_I/2 | =\mathcal {O}(\epsilon)$. Altogether, this implies that the contribution of the imaginary-time contour to the total action is strictly imaginary, and is equal to one-half of the instanton's Euclidean action, up to a factor of ${\cal O}(\epsilon)$. Hence, in the limit $\epsilon \to 0$, we find that the imaginary part of the steadyon's complex action agrees with one-half of the instanton's Euclidean action, such that the exponent of the decay rate satisfies 
\begin{equation}\label{eq:Sconv}
    2 \cdot {\rm Im}(S[\bar{x}]) \to S_E [\bar{x}_{I}].
\end{equation} 
Given our results for the periodic steadyon, it appears reasonable to expect a similar behavior for the steadyon with initial velocity. As shown in Fig.~\ref{fig:SRSEt}, this is indeed the case. To understand this result more generally, we observe that much like the periodic steadyon, the solution with initial velocity can be understood via its analytic continuation evaluated on the complex-time contour $\gamma_\epsilon$. Thus, we may again decompose this complex contour into its projection onto the real- and imaginary-time axes.

\begin{figure}[h!]
    \centering
    \includegraphics[width=0.48\textwidth]{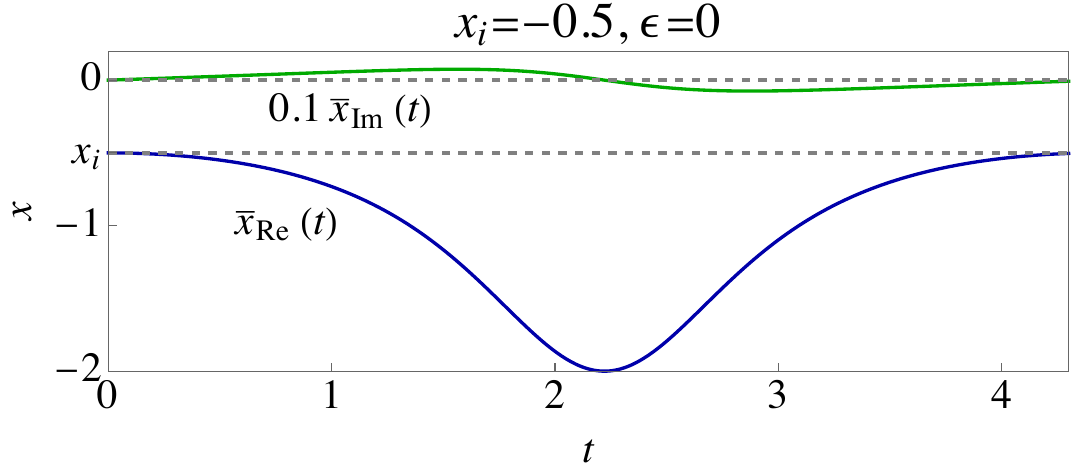}
    \caption{The analytic continuation onto the real-time axis of the steadyon solution with nonvanishing initial velocity which yields the dominant contribution to the overall decay rate.}
    \label{fig:xr}
    \includegraphics[width=0.48\textwidth]{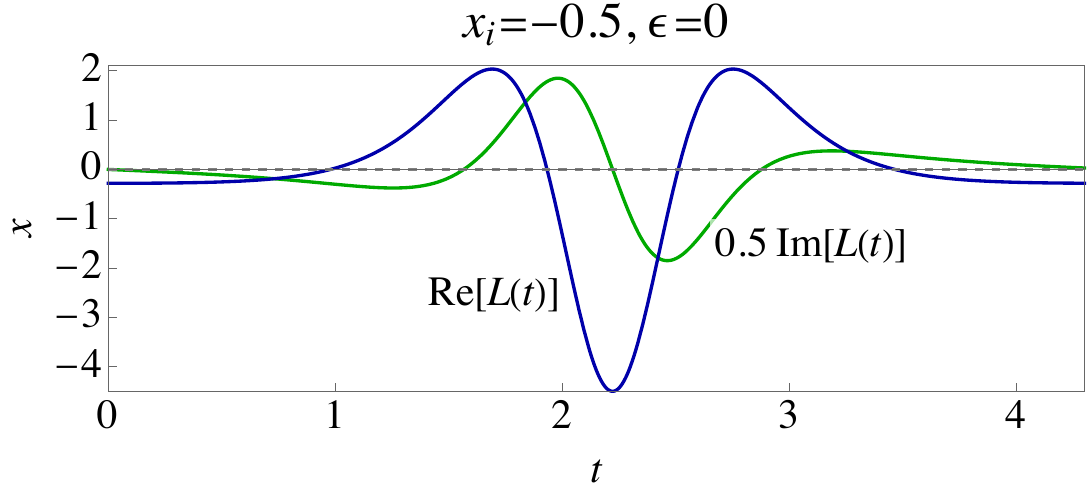}
    \caption{The Lagrange function of the steadyon depicted in Fig.~\ref{fig:xr}. Note that its imaginary part is an anti-symmetric function around the center of the real-time interval of length $\Delta t_R$.}
    \label{fig:Lrt}
\end{figure}

However, \textit{unlike} for the periodic steadyon, we find that the analytic continuation onto the real-time axis is \textit{not} a strictly real function linked to classical behavior, as shown in Fig.~\ref{fig:xr}. This, in turn, yields a complex Lagrange density, see Fig.~\ref{fig:Lrt}. First we note that the double-periodicity of our more general solutions of Eq.~\eqref{sol} manifests, again, on the real-time axis. Next we observe that within each period of length $\Delta t_R$, the Lagrange function is asymmetric, such that its net contribution to the action per interval of this length vanishes. Following our discussion around Eq.~\eqref{Asynch2}, the boundary conditions imply that, in the limit $\epsilon \to \infty$, $t_s \to N \cdot \Delta t_R$, with some positive integer $N$. Thus, although the Lagrange function on the real-time axis is not manifestly real itself, its overall contribution to ${\rm Im} (S [ \bar{x} ])$ vanishes in the appropriate limit. As the contribution from the imaginary-time axis reproduces, by construction, that of the instanton, we recover Eq.~\eqref{eq:Sconv}. Crucially, this discussion is independent of our choice of potential, and in particular does \textit{not} require the potential to have degenerate vacua.

In the imaginary-time picture, it is straightforward to identify the stationary-phase solution corresponding to the dominant contribution to the decay rate. To do so, we solve the equation of motion in imaginary time for varying values of $v_0\equiv \dot{\bar{x}}_{ I} (0)$. This covers, in particular, the bounded solutions giving rise to the solutions of Eq.~\eqref{sol}, as well as unbounded solutions not captured by this expression. Importantly, each of these actions is defined on a different Euclidean-time interval $\Delta \tau_I$, in agreement with our previous discussion.

\begin{figure}[h!]
    \centering
    \includegraphics[width=0.48\textwidth]{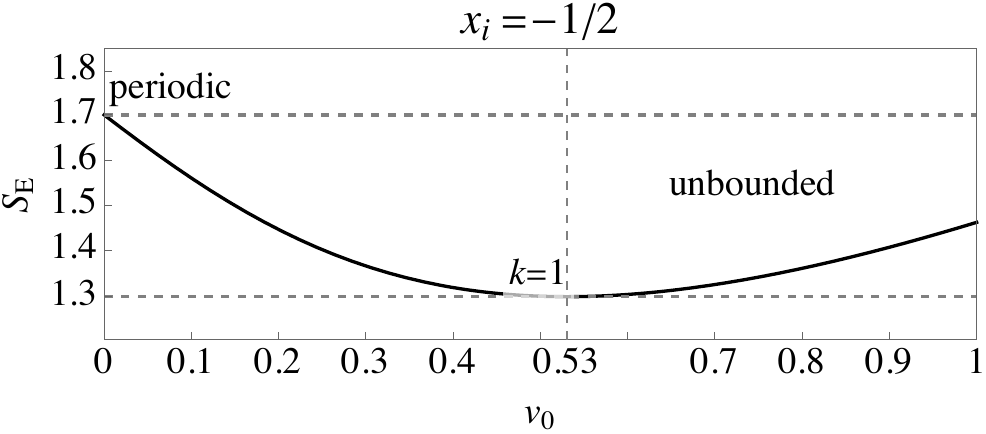}
    \caption{The Euclidean action of the instanton describing tunneling out of our previously considered example. The largest contribution arises from the instanton with initial velocity $v_0 \equiv \dot{\bar{x}}_I (0) \simeq 0.53$, corresponding to the false-vacuum instanton with $k=1$.}
    \label{fig:v0opt}
\end{figure}

In Fig.~\ref{fig:v0opt}, we plot the Euclidean action as a function of $v_0$, again describing the tunneling out of the initial position $x_i=-1/2$, the same example we considered above. 

\section{Normalization factor}\label{sec:Drag}

Having understood the numerator of the decay rate, it remains to evaluate the denominator, i.e., the normalization factor $Z = P_{\Omega_-}$. Once again, we wish to do so by means of the stationary-phase approximation, which requires us to match the solutions of Eq.~\eqref{sol} with the boundary conditions of Eq.~\eqref{eq:zbc}. Before doing so for the real-time case, it is helpful for the interpretation of our results to consider the imaginary-time picture, which we will ultimately recover in the relevant limits.

\subsection{Pulls}

In the limit $x_i \to x_{-}$, the saddle point of the normalization factor is simply given by $\bar{z}_I(\tau) =  x_{-}$. We furthermore anticipate that the saddle-point solution will be defined on an imaginary-time interval of the same length as the instanton, $\Delta \tau_I$. 

\begin{figure}[h!]
    \centering
    \includegraphics[width=0.48\textwidth]{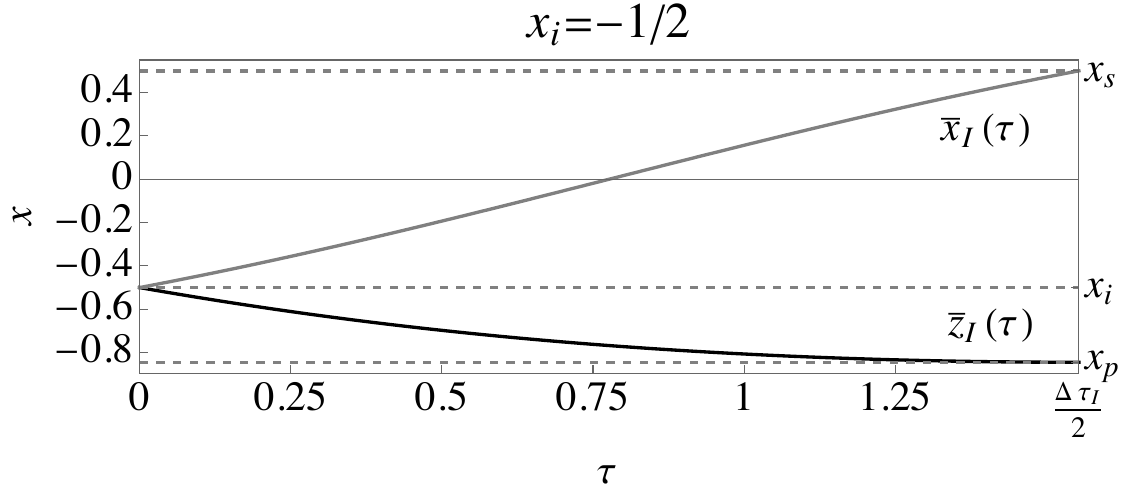}
    \caption{The instanton linked to the dominant contribution to the decay rate, $\bar{x}_I (\tau)$, together with the pull, $\bar{z}_I (\tau)$, for our example with $x_i=-1/2$. }
    \label{fig:Pull}
\end{figure}

This suggests that we consider solutions that begin at $x_i$ with an initial velocity that is just large enough for the particle to reach some turning point $x_p$ at $\tau = \Delta \tau_I/2$ and return to $x_i$ at $\tau = \Delta \tau_I$, as in Fig.~\ref{fig:Pull}. We refer to such solutions as ``pulls'': they begin at $x_i$ and are ``pulled back'' to some point $x_v$ that is also within the false-vacuum basin $\Omega_{-}$. We find that, for any pair $\Delta \tau_I$ and $x_i$, these solutions minimize the Euclidean action, that is, they give rise to the dominant contribution to the normalization factor. This allows us to evaluate the integral over the steadyon's end point $x_v$ in Eq.~\eqref{NormalizationMaster} using a saddle-point approximation around the value of $x_v$ corresponding to this behavior.

For the remainder of this section, we will focus on combinations of $x_i$, $T$ and $\epsilon$ that give rise to the dominant stationary phase, i.e., $k=1$. In particular, these solutions require $\Delta \tau_I \to 0$ as $x_i$ approaches the top of the potential barrier.

\subsection{Drags}

\begin{figure*}[t!]
    \centering
    \includegraphics[width=0.32\textwidth]{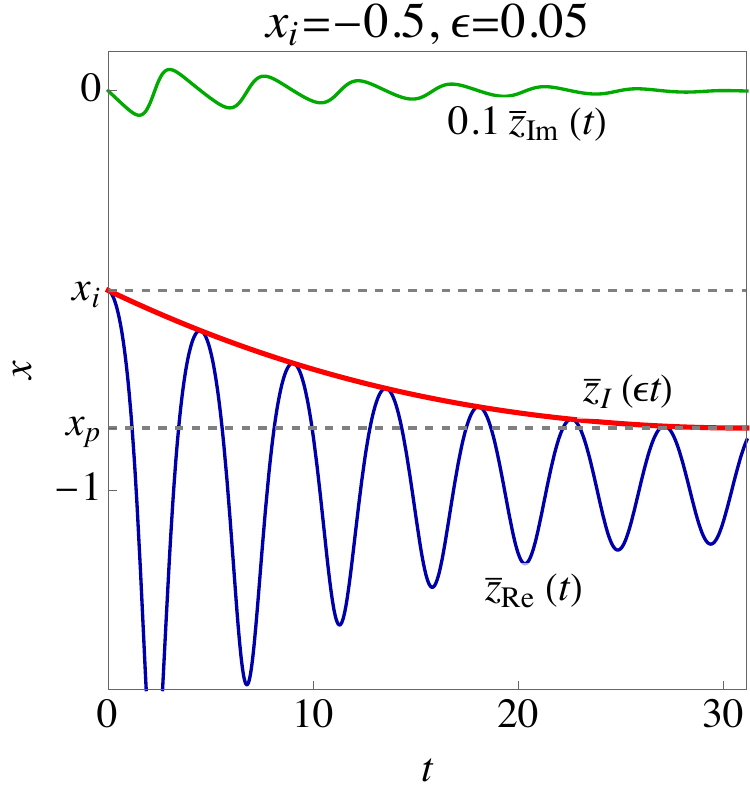}
    \includegraphics[width=0.32\textwidth]{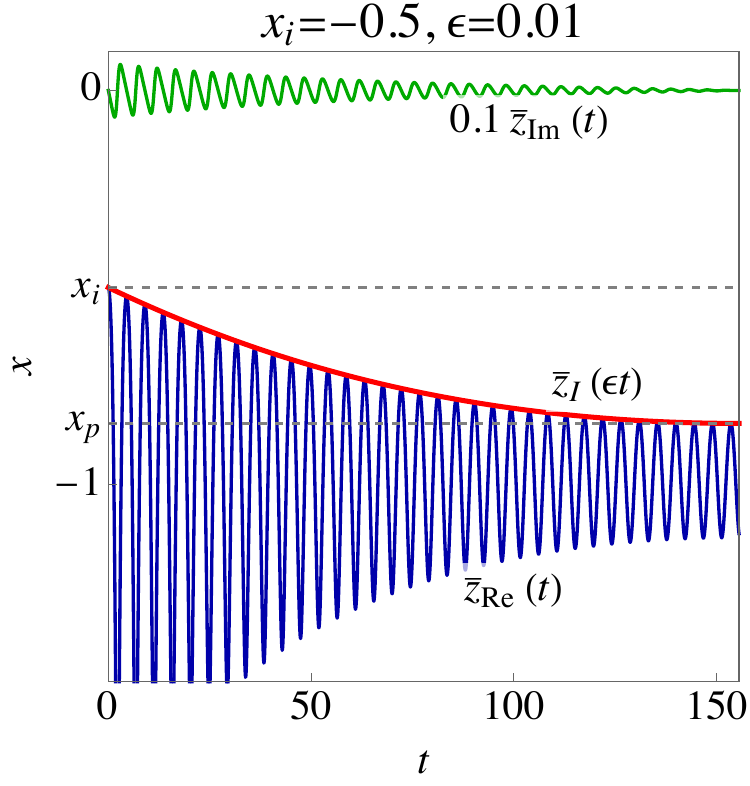}
    \includegraphics[width=0.32\textwidth]{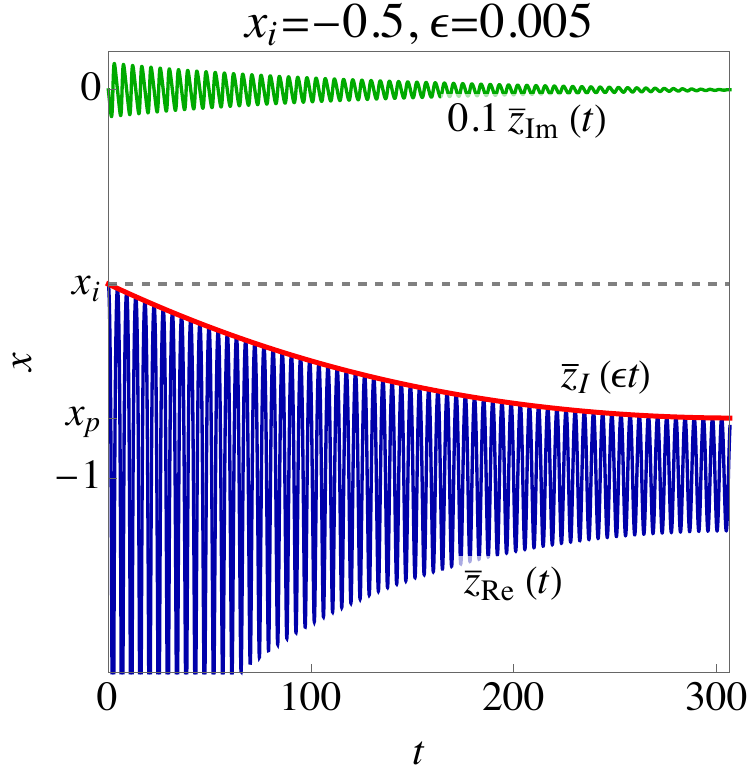}
    \caption{The pull, $\bar{z}_{I}$, and the real and imaginary part of its corresponding drag, $\bar{z}_{\rm Re}$ and $\bar{z}_{\rm Im}$ respectively, for $x_i = - 1/2$}
    \label{fig:ZCI}
\end{figure*}
In order to determine the real-time analogue of the pull---which we dub the ``drag''---we again begin with the general set of solutions given in Eq.~\eqref{sol}. First we recall that the instanton emerged from the time evolution of the amplitude of the real part of the steadyon solution. This suggests that we consider the complex solution of Eq.~\eqref{sol}, which similarly reproduces the pull, as shown in Fig.~\ref{fig:ZCI}. Crucially, the parameters of this solution are the same as those of the corresponding pull. In other words, $\epsilon$ is fixed by the requirement that the steadyon solution exists, while $k$ and $b$ are determined by $x_i$ and $T$. 

The drag can also be understood along the lines of our discussion in Secs.~\ref{sec:SteadyonSolution} and~\ref{sec:NPSteadyonSolution}. At $t=0$, the imaginary part of the particle's motion begins with a large velocity dampening the oscillatory motion of the real part within the false-vacuum basin. At $t=t_s$, the amplitude of the latter reaches its smallest value within this process, which is given by the turning point of the pull. 

Unlike the steadyon solution, the drag is not affected by the small but finite mismatch between the analytic solution and the relevant boundary conditions, as discussed (for the steadyon case) below Eqs.~(\ref{Asynch1})--(\ref{Asynch2}). First, unlike for the steadyon, the value $x_v$ of the drag at $t=t_s$ is not fixed, but can take any \textit{real} value, since in the end we perform the remaining integral over $x_v$. The boundary condition for the imaginary part, meanwhile, can be satisfied in the limit $\epsilon \to 0$ by choosing the elliptic modulus $k$ corresponding to the pull, as shown in Fig.~\ref{fig:ZCI}. The fact that the pull reaches its turning point at $\tau \simeq t_s \cdot \epsilon$ translates to a roughly constant oscillation amplitude for the real part of the drag, namely, an oscillation of the particle around the false vacuum. In the limit $\epsilon \to 0$, this approximates a classically allowed motion, and hence $\bar{z}_{\rm Im} (t_s) \to 0$ in the limit $\epsilon \to 0$. Thus, the boundary conditions of Eq.~\eqref{eq:zbc} are indeed satisfied by the drag with the choice of $k$ corresponding to the pull in the limit $\epsilon \to 0$.\footnote{Alternatively, this allows one to understand the choice of $k$ as a consequence of the requirement that the imaginary part of the steadyon vanishes at the emergence time $t=t_s$.}

\subsection{Real-time action and pull limit}

\begin{figure}[h!]
    \centering
    \includegraphics[width=0.48\textwidth]{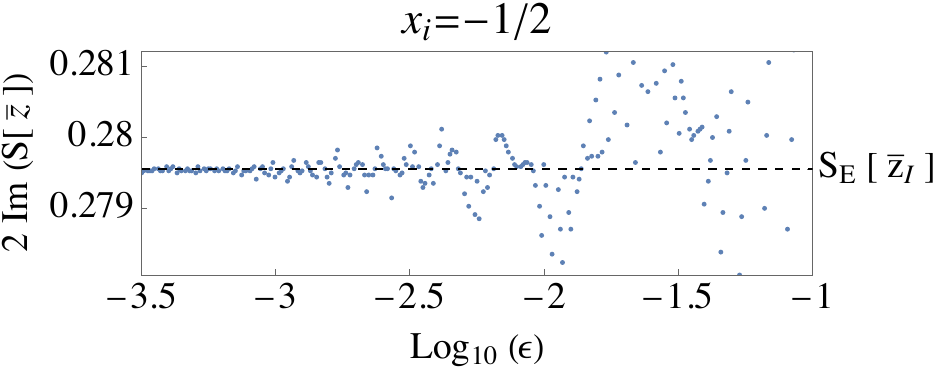}
    \caption{The imaginary part of the exponent of the normalization factor. We again recover the desired Euclidean-time result in the limit $\epsilon \to 0$.}
    \label{fig:SRSEz}
\end{figure}

In Sec.~\ref{sec:SteadyonAction}, we demonstrated that the imaginary part of the action for both periodic and non-periodic steadyon solutions converges to the Euclidean action of the associated instanton in the appropriate limit, $\epsilon \to 0, \, T\to \infty$. As shown explicitly in Fig.~\ref{fig:SRSEz}, this is also the case for the drag $\bar{z} (t)$ and its corresponding pull $\bar{z}_I (\tau)$. Just as for the steadyon, this behavior can be understood by considering the analytic continuation of $\bar{z} (t)$ onto the full complex-time plane. 

In Sec.~\ref{sec:FirstPrinciples}, we had found that the numerator of the decay rate is just the time-derivative of its denominator. This suggests that one use the same combination $(\epsilon, t_s)$ for the drag and steadyon, with the latter establishing a dependence $t_s (\epsilon)$. Thus, the action of the drag can be understood as an integral over the same complex-time contour $\gamma_\epsilon$ as the steadyon, which we can once again decompose into its projections onto the real- and imaginary-time axes. Just as for the steadyon solution with initial velocity, we find that the drag's projection onto the real-time axes does \textit{not} describe a classically allowed motion. This gives rise to a complex Lagrange function, which is again anti-symmetric within each interval of length $\Delta t_R$. 

However, since the drag $\bar{z} (t)$ corresponds to a different choice of the elliptic modulus $k$ than for the steadyon solution $\bar{x} (t)$, the period of the drag's real-time projection does not necessarily match that of the steadyon. Thus, in general, the value of the drag at $t=t_s$ is not necessarily identical to the turning point $x_p$ of the pull. As the latter serves as the initial value for dynamics along the imaginary-time axis, the projection of the drag onto the imaginary-time axes does not necessarily reproduce the pull, and the integral over the real-time axis contributes to the imaginary part. The results shown in Fig.~\ref{fig:SRSEz} suggest that the latter compensates precisely for the deviation in the contribution from the imaginary-time axis. 

\begin{figure*}[t!]
    \centering
    \includegraphics[width=0.32\textwidth]{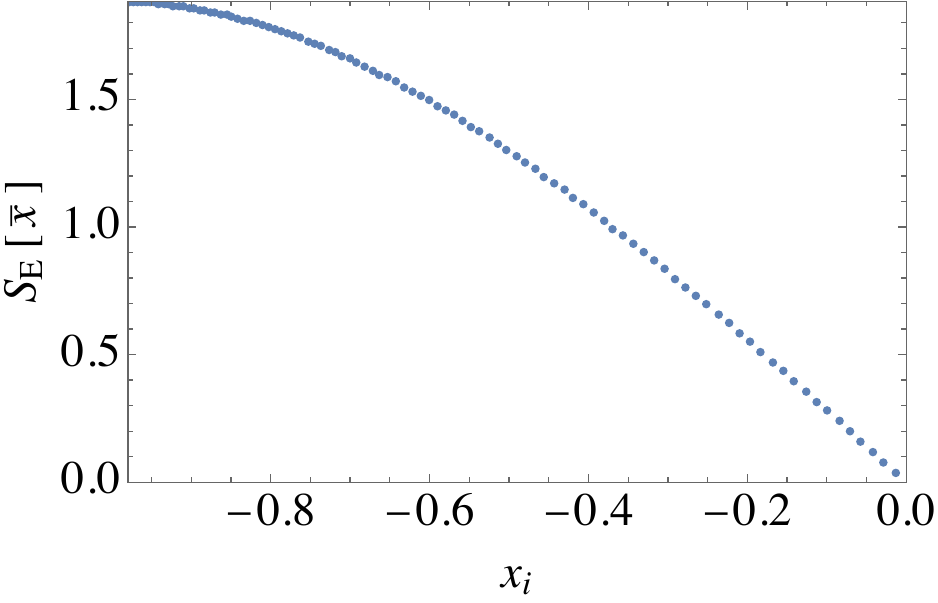}
    \includegraphics[width=0.32\textwidth]{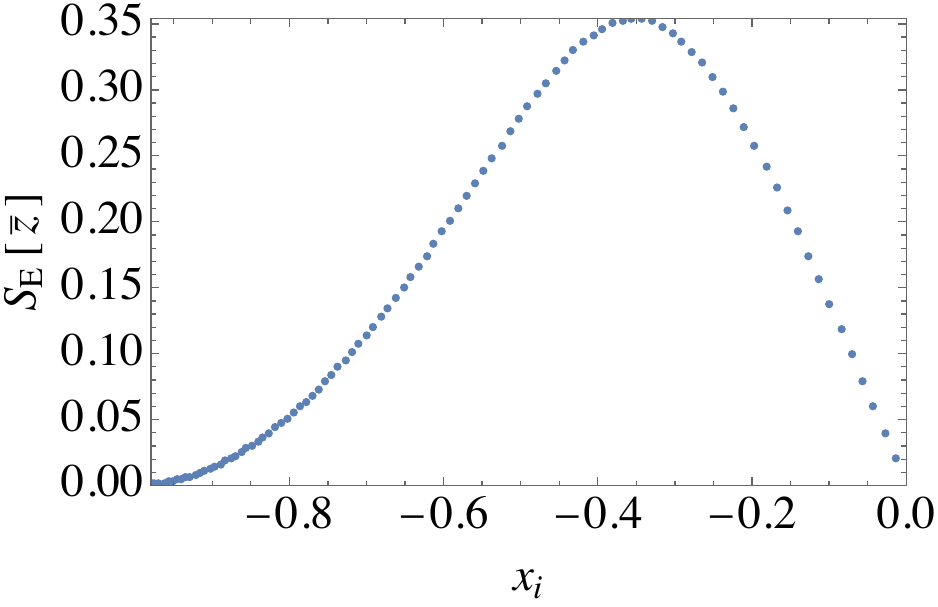}
    \includegraphics[width=0.32\textwidth]{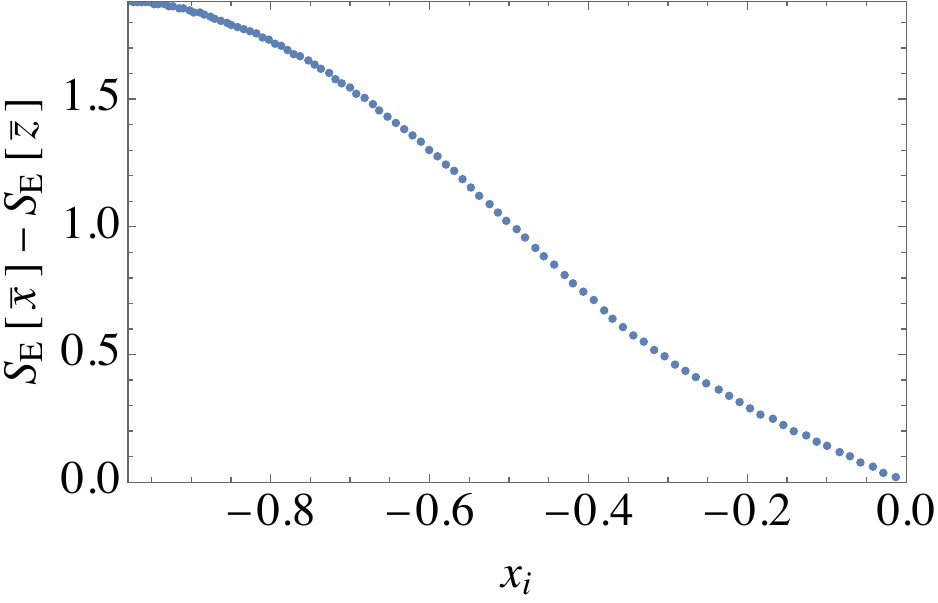}
    \caption{\textit{Left panel:} The Euclidean action of the dominant instanton as a function of the initial positon $x_i$. \textit{Middle panel:} The Euclidean action of the pull corresponding to the dominant instanton as a function of the initial positon $x_i$. \textit{Right panel:} The leading order exponent of the decay rate as a function of the initial positon $x_i$. The initial position $x_1=-1$ corresponds to tunneling out of the false vacuum. In this limit, the Euclidean action of the pull $S_E [\bar{z}]$ vanishes and that of the instanton, $S_E [\bar{x}]$, reproduces that of the familiar bounce.}
    \label{fig:DR}
\end{figure*}

Our previous reasoning can now also be used to quickly understand this behavior. Whereas in general $x_v \neq x_p$, it is easy to see that for any $x_i$ there exists a (countable) infinite number of choices for $\epsilon$ and $t_s$ for which $x_v \simeq x_p$ to arbitrary accuracy. For these, the integral over the real-time axis contains a whole number of periods, each of which gives rise to a strictly real action along the real-time axis. Having $x_v \simeq x_p$ also implies that the solution along the imaginary-time contour is, to very good accuracy, given by the pull, such that the contribution from this branch is nothing but the Euclidean action of the pull solution. As we can expect the path integral to converge to a well-defined value, we can use this countable set of stationary-phase results to deduce this value---and thus, by extension, also the value of the complex-valued action for those values of $\epsilon$ for which the saddle point exists, but for which the action takes a less straightforward form.

\subsection{Effect of the pull on the leading-order decay rate}

We can distinguish two regimes for the pull, based on the value of $x_p$. First, if $x_i$ is sufficiently close to $x_{-}$ and $\Delta \tau_I$ is large enough, the Euclidean action is minimized for values of $x_p$ close to $x_{-}$. This implies that, for most of the time interval $\Delta \tau_I$, the particle hovers near $x_{-}$, corresponding to a near-vanishing Euclidean Lagrangian. In order to get to this point, the particle needs a large enough initial velocity, giving rise to a contribution to the Euclidean action from the kinetic term. If, on the other hand, $x_i$ is further away from $x_{-}$ and $\Delta \tau_I$ is small enough, the reduction in Euclidean action achieved by approaching $x_p$ cannot compensate for this additional contribution. In the most extreme case of a near-vanishing $\Delta \tau_I$, this can lead to the pull essentially approaching a constant function with $x_p \simeq x_i$, with a vanishing Euclidean action. 

We find that this last scenario is indeed realized in our toy-model, with $\Delta \tau_I \to 0$ as $x_i$ approaches the top of the potential barrier at $x_i = 0$. See Fig.~\ref{fig:DR}, in which we present the Euclidean actions of the dominant instanton, its corresponding pull, and the overall leading-order exponent of the tunneling rate. First, we observe that tunneling becomes unsuppressed as we approach the top of the potential barrier ($S_E [ \bar{x} ] \to 0$), as expected. Next, we find that the normalization factor gives rise to a significant contribution to the tunneling rate in the intermediate regime.

\section{Conclusion}\label{sec:Conclusion}

We have studied the calculation of tunneling rates from first principles using real-time techniques. On a technical level, our most important result is a strategy for the evaluation of the path integrals expected to appear in such calculations. To avoid the complications linked to a Wick rotation by a finite angle, we regularize the path integral by introducing a small imaginary part for the Hamiltonian, controlled by a regularization parameter $\epsilon$. This procedure is well-defined in both quantum mechanics and quantum field theory, and has also been shown to produce meaningful results in curved backgrounds and even for systems with an explicit time-dependence~\cite{Kaya:2018jdo}.

For systems without an explicit time-dependence, this step is equivalent to an infinitesimal Wick rotation onto a complex-time contour $\gamma_\epsilon$. We find that for any physical time $T$ subject to the usual assumptions, there exist choices of $\epsilon$ representing a one-parameter family of approximate saddle points. Crucially, in the limit $T\to \infty$, the existence of these saddle points requires $\epsilon \to 0$, consistent with the role of $\epsilon$ as a regulator. For our (unphysical) choice of initial state, we find an additional condition on the physical times for which these saddle points exist, in agreement with the interpretation of the state.

We find that, for systems without an explicit time-dependence, the relevant limits reproduce the familiar result obtained in the usual imaginary-time picture. This requires the steadyon's analytic continuation to be doubly-periodic and regular between the real-time axis and $\gamma_\epsilon$. This can be expected for a wide class of physical systems and is easily confirmed for the imaginary-time instanton, allowing for a straightforward generalization of our result.

Our discussion focuses primarily on tunneling within a given potential for arbitrary initial positions. One could deploy the formalism we develop to further study how the tunneling rate would change as one modifies the potential as well. Our analysis strongly suggests that the effects of varying the height of the potential barrier would affect the final results in a way similar to what one would expect from the WKB approximation. Overall, we have developed a general formalism with which one could vary both the initial conditions and the particular shape of the potential.

It is worth noting that these properties are, in general, non-trivial. For field-theoretical systems, for example, dissipation effects can prevent the existence of a periodic solution, e.g., if the initial state of interest is some localized excitation, or if one is considering a system within a time-dependent background. The complications linked to these properties, however, would only affect the instanton limit---in other words, our general strategy for calculating the decay rate using steadyons remains valid for such systems, whereas the usual Wick-rotated instantons do not.

While motivated by the need to realize the boundary conditions describing quantum tunneling, our method can also be used to evaluate more general path integrals with non-trivial boundary conditions. This includes, in particular, the normalization factor of the tunneling rate. In the real-time picture, the latter is dominated by solutions which we called \textit{drags}, which we have shown to be linked to imaginary-time solutions called \textit{pulls} in a similar way as the steadyons are related to instantons. The ``drag'' contributions stem from summing over paths that remain within the false-vacuum basin $\Omega_{-}$.

Our work leaves open two important conceptual questions. First, it appears evident that our technique can be readily generalized to more complicated systems, in particular those with an explicit time-dependence. As our recovery of the imaginary-time picture hinged on the absence of such a dependence, considering such systems offers a promising path towards an even deeper understanding of quantum tunneling phenomena across a broad range of scenarios. 

Second, our analysis so far captures only the leading-order contribution to the decay rate. It is, however, well-known that many important aspects of the tunneling rate can only be understood at the next-to-leading order. This includes the prefactor of the exponent---that is, the term $A$ in Eq.~(\ref{eq:GammaUs})---as well as the effect of possible zero modes. For the tunneling rate out of a false vacuum, it is well-known that eliminating the latter introduces an additional imaginary part to the decay rate, which, in some cases, can exactly cancel the contributions of some individual stationary phases~\cite{Andreassen:2016cvx}. While this task would be straightforward once one has transitioned to the imaginary-time picture, it has recently been shown that this transition itself is already quite non-trivial for the simpler case of tunneling out of a false vacuum~\cite{Ai:2019fri}.\footnote{For a discussion of the functional determinants of periodic instantons in this case, consider, e.g., Refs.~\cite{Liang:1992ms,Liang:1994xn}. A similar discussion can be found in Ref.~\cite{Gildener:1977sm}. Most important, these authors address the zero mode by introducing a time-dependent collective coordinate, which also allows for the preservation of the stricter boundary conditions emerging from our approach. For a more general discussion of such degrees of freedom, consider Ref.~\cite{Steingasser:2020lhj}.}

\vskip 30pt
\begin{acknowledgments}\ignorespaces
We thank Wen-Yuan Ai, Morgane K\"{o}nig, Bruno Scheihing and Matthew D. Schwartz for valuable discussions, as well as Nils Wagner for pointing out several possibly unclear formulations in an earlier version. TS's contributions to this work were made possible by the Walter Benjamin Programme of the Deutsche Forschungsgemeinschaft (DFG, German Research Foundation) -- 512630918. Portions of this work were conducted in MIT's Center for Theoretical Physics and partially supported by the U.S. Department of Energy under Contract No.~DE-SC0012567. This project was also supported in part by the Black Hole Initiative at Harvard University, with support from the Gordon and Betty Moore Foundation and the John Templeton Foundation. 

The opinions expressed in this publication are those of the author(s) and do not necessarily reflect the views of these Foundations.
\end{acknowledgments}

\bibliography{ref}

\end{document}